\documentclass[12pt]{iopart}

\usepackage{graphicx}

\usepackage{dcolumn,multirow,color}

\begin{document}

\title[The position profiles of order cancellations]{The position profiles of order cancellations in an emerging stock market}

\author{Gao-Feng Gu$^{1,2}$, Xiong Xiong$^{3}$, Fei Ren$^{1,2,4}$, Wei-Xing Zhou$^{1,2,4}$ and Wei Zhang$^{3}$}
\address{$^1$ School of Business, East China University of Science and Technology, Shanghai 200237, China}
\address{$^2$ Research Center for Econophysics, East China University of Science and Technology, Shanghai 200237, China}
\address{$^3$ College of Management and Economics, Tianjin University, Tianjin 300072, China}
\address{$^4$ School of Science, East China University of Science and Technology, Shanghai 200237, China}
\ead{wxzhou@ecust.edu.cn; weiz@tju.edu.cn}

\begin{abstract}
  Order submission and cancellation are two constituent actions of stock trading behaviors in order-driven markets. Order submission dynamics has been extensively studied for different markets, while order cancellation dynamics is less understood. There are two positions associated with a cancellation, that is, the price level in the limit-order book (LOB) and the position in the queue at each price level. We study the profiles of these two order cancellation positions through rebuilding the limit-order book using the order flow data of 23 liquid stocks traded on the Shenzhen Stock Exchange in the year 2003. We find that the profiles of relative price levels where cancellations occur obey a log-normal distribution. After normalizing the relative price level by removing the factor of order numbers stored at the price level, we find that the profiles exhibit a power-law scaling behavior on the right tails for both buy and sell orders. When focusing on the order cancellation positions in the queue at each price level, we find that the profiles increase rapidly in the front of the queue, and then fluctuate around a constant value till the end of the queue. These profiles are similar for different stocks. In addition, the profiles of cancellation positions can be fitted by an exponent function for both buy and sell orders. These two kinds of cancellation profiles seem universal for different stocks investigated and exhibit minor asymmetry between buy and sell orders. Our empirical findings shed new light on the order cancellation dynamics and pose constraints on the construction of order-driven stock market models.

\vspace{2pc}
\end{abstract}



\maketitle

\section{Introduction}
\label{sec:introduction}

In order-driven markets, order cancellation plays an important role in the price formation and makes significant contribution to the order book activity. When limit orders at the best price cancelled completely, mid-price (the mean of best bid and best ask) and spread will change as well. If cancellation occurs inside the limit-order book (LOB), it also affects the shape of LOB and has potential effects on price formation.

The motivation of order cancellation is to avoid risk including non-execution (NE) risk and free option (FO) risk. The former risk arises when the current price moves away from the submitting price, so limit orders may not to be executed immediately. NE risk may cause traders to suffer the opportunity cost \cite{Griffiths-Smith-Turnbull-White-2000-JFE}. In order to avoid NE risk, traders may submit more aggressive limit orders or market orders to increase the execution probability. FO risk arises when good or bad news arrives, which drives the intrinsic value of asset underestimated or overestimated based on the current market price. Traders cancel their stale orders in time to prevent them from being traded at unfavorable prices. Thus, if traders concern the NE risk, they will revise their orders to increase the price priority, while if traders focus on the FO risk, they will revise orders to decrease the price priority. In the Australian security market, NE risk is the major reason for order cancellation, especially for large orders \cite{Fong-Liu-2010-JBF}.

With the reason that order cancellation concerns the dynamics of LOB and there are not enough cancellation data flow in the past, only a few papers refer to the study of order cancellation. However, in recent years, with the development of information technology and rapid development of electronic trading venues, traders can easily make an order cancellation at their private computers, which causes a striking increase of order cancellation in financial markets. For example, more than 20\% of submitted orders are cancellation orders, and 40\% of limit orders are cancelled on the New York Stock Exchange \cite{Yeo-2005-XXXX}. The cancellation percentage is even higher in futures markets \cite{Coppejans-Domowitz-2002-XXXX}, where the cancellation rate is as high as 68.3\%.

Many scholars have paid attention to the study of order cancellation in the recent years and presented useful results. Yeo \cite{Yeo-2005-XXXX} studied the database of 148 stocks traded on the New York Stock Exchange in the year 2001 and found that about 95\% order cancellations occur within 10 minutes after submission. He also reported that traders tend to resubmit limit orders with more aggressive prices after a cancellation to avoid NE risk. Fong and Liu \cite{Fong-Liu-2010-JBF} studied the order flow data of 40 stocks listed on the Australian Securities Exchange (ASX) in August 2000. They found that half cancellations occur within the first two price levels in the LOB, and indicated that cancellation probability increases with order size, but decreases with order aggressiveness. Liu \cite{Liu-2009-JFM} presented a simple model for order revision and cancellation. The paper shows that cancellation is positively related to order submission risks, and that large capitalization stocks tend to have more cancellations. Hasbrouck and Saar \cite{Hasbrouck-Saar-2002-XXXX} focused on the cancellations on the Island electronic communications network (ECN). They observed that a large number of limit orders are cancelled very shortly after their submissions and two sharp jumps appear in the probability density function (PDF) of order cancellation. Biais {\em et al}. \cite{Biais-Hillion-Spatt-1995-JF} analyzed the database of 40 stocks in the year 1991 in the Paris Bourse, and found that cancellations at the bid or ask side are relatively frequent after large sales or purchases. The paper shows that cancellations follow each other quickly on the same side of the LOB as well. Ellul {\em et al}. \cite{Ellul-Holden-Jain-Jennings-2003-XXXX,Ellul-Holden-Jain-Jennings-2007-JEF} studied the data sample of 148 stocks trading on the NYSE, and found that when a buy or sell order is cancelled, the most likely subsequent event is the arrival of the same kind of limit orders. Ni  {\em et al}. \cite{Ni-Jiang-Gu-Ren-Chen-Zhou-2010-PA} analyzed the statistical properties of inter-cancellation duration (defined as the waiting time between consecutive cancellations) of 22 stocks in the Chinese stock market. They found that the probability density function of durations can be fitted by a Weibull distribution, and that the duration time series exhibits long-term memory and multifractal nature with the detrended fluctuation analysis (DFA) and the multifractal DFA methods. They made a conclusion that the order cancellation is a non-Poisson process.

The relationship between cancellation and bid-ask spread has been wildly studied, while the results are controversial. Some scholars concluded that cancellation is positively related to spread \cite{Fong-Liu-2010-JBF,Yeo-2005-XXXX}, others however showed that cancellation is more likely to happen when spread is tight \cite{Liu-2009-JFM,Ellul-Holden-Jain-Jennings-2003-XXXX,Ellul-Holden-Jain-Jennings-2007-JEF}. In addition, Order cancellation is more frequent at the opening and closing time on a trading day, displaying a U-shape intra-day pattern \cite{Fong-Liu-2010-JBF,Coppejans-Domowitz-2002-XXXX,Liu-2009-JFM,Biais-Hillion-Spatt-1995-JF,Ellul-Holden-Jain-Jennings-2007-JEF,Ellul-Holden-Jain-Jennings-2003-XXXX}.

In the paper, we will study the empirical position profiles of order cancellations by rebuilding the LOBs using 23 liquid
stocks traded on the Shenzhen Stock Exchange (SZSE) in the Chinese stock market. The rest of paper is organized as follows. In Section~\ref{sec:database}, we will describe briefly the database we adopt. Section~\ref{sec:PDF-LOB} presents the cancellation position profile of price levels in the LOB. We further discuss the cancellation position in a queue at a certain price level and study its profile in Section~\ref{sec:PDF-PL}. Section~\ref{sec:conclusion} summarizes the results.

\section{Data sets}
\label{sec:database}

There are two kinds of auctions held on the Shenzhen Stock Exchange, called opening call auction and continuous auction. Opening call auction is held to generate the opening price at the beginning of a trading day. It corresponds to the process of one-time centralized matching of buy and sell orders accepted during the period from 9:15 am and 9:25 am. Continuous auction operates from 9:30 am to 11:30 am and 13:00 pm to 15:00 pm. It refers to the process of continuous matching of buy and sell orders on a one-by-one basis. The interval between opening call auction and continuous auction (9:25 am - 9:30 am) is the cool period, when the Exchange is open to orders routing from members, but does not process orders or process cancellations. Closing call auction (14:57 pm - 15:00 pm) was not held until July 1 2006 to generate the closing price.

In an order-driven market, Limit-order book is a queue of limit orders waiting to be executed and it is the base of continuous double auction mechanism. Limit orders are arranged in the LOB according to the price-time priority which consists of price priority principle and time priority principle. Price priority principle means priority is given to a higher buy order over a lower buy order and a lower sell order is prioritized over a higher sell order. Time priority principle concerns the orders having the same submitting price and indicates that the priority is given to the order earlier received by the exchange trading system. Limit-order book is constructed by the price-time priority principle which consists of price priority principle and time priority principle.

\begin{figure}[htb]
\centering
\includegraphics[width=12cm]{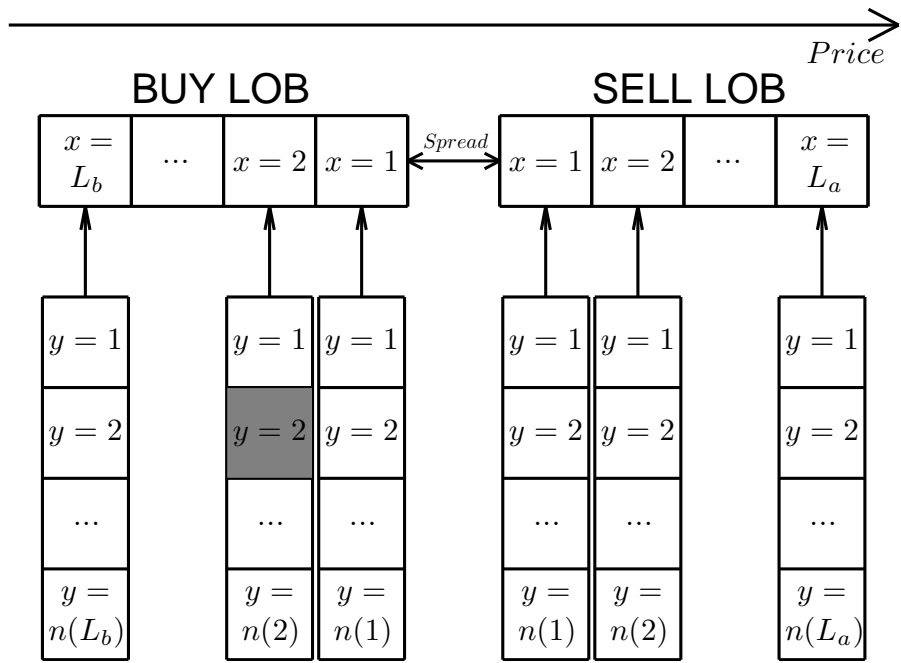}
\caption{\label{Fig:LOB} The structure of limit order book.}
\end{figure}

Price levels in the LOB are discrete. The difference between two adjacent price levels is the tick size or its multiple. In the Chinese stock market, tick size is 0.01 CNY for all the A shares of stocks. Figure~\ref{Fig:LOB} presents the structure of LOB. In the figure, we denote $x$ as the price level in the LOB. In the buy LOB, orders with higher limit prices have executing priority and are stored in the front of the LOB, so $x$ decreases with buy price. In contrast, in the sell LOB, priority is given to the orders with lower limit prices, and $x$ increases with the sell price. The highest buy price allocated at the first price level in the buy LOB ($x=1$) is called the best bid, and the lowest sell price allocated at the first price level in the sell LOB ($x=1$) is called the best ask. Spread is defined as the difference between the best bid and best ask. When limit orders are submitted with the same price, they are also stored in a queue at a certain price level according to the arriving time. Denote $y$ as the sequence of orders' arriving time, and the smaller value of $y$ indicates the earlier arriving time among the orders with the same submitting price.

Our study is based on the order flow data of 23 liquid stocks extracted from the A-share market on the SZSE in 2003. The database contains details of order placement and order cancellation with the time stamp accurate to 0.01 second. Table~\ref{Tb:Num_C} depicts the cancellation number $C$ for both buy and sell orders of 23 stocks. We calculate the ratio $r$ of the number of cancelled buy (or sell) orders to the number of all buy (or sell) orders including effective limit orders placed in the LOB and marketable orders that are fully or partly filled. We find that the ratio $r$ fluctuates within a wide range, with $[4.2\%,~24.7\%]$ for buy orders and with $[4.5\%,~24.3\%]$ for sell orders. There is an interesting feature that the ratio of buy orders is close to sell orders for each stock, which means that a large proportion of buy orders cancelled corresponds to a large number of cancellations taking place for sell orders, and vice versa. For all the stocks, we have the mean values $\overline{r}=16.1\%$ for buy orders and $\overline{r}=17.1\%$ for sell orders.

\begin{table}[htp]
  \centering
  \caption{The number and percentage of cancellations of buy and sell orders in the continuous auction for 23 stocks. The columns show the stock code, the number of cancellations $C$, the ratio $r$ of the number of cancelled buy (or sell) orders to the number of all buy (or sell) orders, respectively.}
  \medskip
  \label{Tb:Num_C}
  \centering
  \begin{tabular}{crrrrc|ccrrrr}
  \hline \hline
  \multirow{3}*[2mm]{Stock} & \multicolumn{2}{@{\extracolsep\fill}c}{Buy orders} & \multicolumn{2}{@{\extracolsep\fill}c}{Sell orders} &&& \multirow{3}*[2mm]{Stock} & \multicolumn{2}{@{\extracolsep\fill}c}{Buy orders} & \multicolumn{2}{@{\extracolsep\fill}c}{Sell orders} \\
  \cline{2-3} \cline{4-5} \cline{9-10} \cline{11-12}
  & \multicolumn{1}{@{\extracolsep\fill}c}{$C$} & \multicolumn{1}{@{\extracolsep\fill}c}{$r$(\%)} & \multicolumn{1}{@{\extracolsep\fill}c}{$C$} & \multicolumn{1}{@{\extracolsep\fill}c}{$r$(\%)} &&&& \multicolumn{1}{@{\extracolsep\fill}c}{$C$} & \multicolumn{1}{@{\extracolsep\fill}c}{$r$(\%)} & \multicolumn{1}{@{\extracolsep\fill}c}{$C$}& \multicolumn{1}{@{\extracolsep\fill}c}{$r$(\%)} \\
  \hline
    000001 & 317,015 & 19.2 & 274,929 & 18.3 &&& 000406 & 94,100  & 20.3 & 91,776  & 20.1 \\
    000002 & 34,577  &  4.2 & 43,801  &  4.9 &&& 000429 & 36,999  & 20.1 & 36,174  & 17.4 \\
    000009 & 183,804 & 21.9 & 187,421 & 20.7 &&& 000488 & 32,585  & 18.2 & 33,854  & 18.4 \\
    000012 & 114,662 & 24.7 & 106,361 & 24.3 &&& 000539 & 26,950  & 15.0 & 27,087  & 15.9 \\
    000016 & 60,219  & 21.1 & 59,189  & 19.2 &&& 000541 & 19,715  & 18.8 & 19,847  & 16.6 \\
    000021 & 157,174 & 23.8 & 152,853 & 22.0 &&& 000550 & 122,865 & 23.6 & 129,606 & 22.9 \\
    000024 & 42,593  & 22.3 & 45,594  & 19.6 &&& 000581 & 27,236  & 18.6 & 29,490  & 16.2 \\
    000027 & 33,058  &  7.1 & 38,647  &  5.8 &&& 000625 & 123,361 & 22.2 & 131,719 & 22.6 \\
    000063 & 25,681  &  7.2 & 34,815  &  7.5 &&& 000709 & 65,704  & 18.7 & 64,017  & 17.1 \\
    000066 & 110,289 & 24.3 & 106,695 & 21.8 &&& 000720 & 16,558  & 11.4 & 14,209  & 10.7 \\
    000088 & 6,861   &  5.6 & 6,917   &  4.5 &&& 000778 & 43,576  & 18.4 & 47,360  & 16.4 \\
    000089 & 20,909  &  7.0 & 22,984  &  7.6 &&& \\
  \hline\hline
 \end{tabular}
\end{table}

The cancellation behavior is related to the aggressiveness of submitted orders. There are five types of orders according to their submission aggressiveness, that is, fully filled orders, partially filled orders, orders placed inside the bid-ask spread, orders placed at the same best price, and orders placed inside the LOB. For buy or sell orders of each type except for fully filled orders that do not have cancellations, we calculate the ratio ($r_1$, $r_2$, $r_3$, and $r_4$) of the number cancellations to the number of all order of the same type and order direction. The result is presented in table~\ref{Tb:Ps_C}.

\begin{table}[htp]
  \centering
  \caption{Percentages of cancellations of buy (or sell) orders of the four types for the 23 stocks investigated. The columns show the stock code, the ratio $r_1$ of the number of cancellations of partially filled buy (or sell) orders to the total number of partially filled buy (or sell) orders, the ratio $r_2$ of the number of cancellations of buy (or sell) orders placed inside the spread to the total number of buy (or sell) orders placed inside the spread, the ratio $r_3$ of the number of cancellations of buy (or sell) orders placed at the same best price to the total number of buy (or sell) orders placed at the same best price, and the ratio $r_4$ of the number of cancellations of buy (or sell) orders placed inside the book to the total number of buy (or sell) orders placed inside the book.}
\medskip
\label{Tb:Ps_C}
\centering
\begin{tabular}{ccrrrrcrrrr}
 \hline \hline
\multirow{3}*[2mm]{Stock} && \multicolumn{4}{@{\extracolsep\fill}c}{Buy orders (\%)} && \multicolumn{4}{@{\extracolsep\fill}c}{Sell orders(\%)} \\ \cline{3-6} \cline{8-11}
 && \multicolumn{1}{@{\extracolsep\fill}c}{$r_1$} & \multicolumn{1}{@{\extracolsep\fill}c}{$r_2$} &
\multicolumn{1}{@{\extracolsep\fill}c}{$r_3$} &
\multicolumn{1}{@{\extracolsep\fill}c}{$r_4$}&& \multicolumn{1}{@{\extracolsep\fill}c}{$r_1$} & \multicolumn{1}{@{\extracolsep\fill}c}{$r_2$} &
\multicolumn{1}{@{\extracolsep\fill}c}{$r_3$} & \multicolumn{1}{@{\extracolsep\fill}c}{$r_4$} \\
  \hline
    000001 && 0.6 & 10.0 & 16.1 & 30.5 && 0.5 &  3.5 & 18.9 & 25.7 \\
    000002 && 0.1 &  2.2 &  4.8 &  6.3 && 0.1 &  2.1 &  6.0 &  6.5 \\
    000009 && 0.6 & 13.8 & 20.2 & 34.7 && 0.5 &  4.5 & 22.6 & 26.9 \\
    000012 && 0.6 & 11.6 & 25.3 & 41.0 && 0.7 & 14.4 & 28.8 & 35.4 \\
    000016 && 0.5 & 13.6 & 22.3 & 34.1 && 0.4 & 12.9 & 23.9 & 26.4 \\
    000021 && 0.6 & 13.2 & 24.9 & 38.0 && 0.5 & 11.1 & 26.9 & 30.4 \\
    000024 && 0.4 & 14.9 & 24.5 & 35.1 && 0.4 & 17.2 & 23.6 & 26.0 \\
    000027 && 0.4 &  4.3 &  7.4 & 10.5 && 0.2 &  3.6 &  6.7 &  7.4 \\
    000063 && 0.4 &  6.1 &  9.2 & 11.2 && 0.7 &  6.9 &  9.7 &  9.5 \\
    000066 && 0.8 & 15.4 & 25.1 & 38.5 && 0.8 & 10.5 & 26.8 & 29.6 \\
    000088 && 0.3 &  7.5 &  8.3 &  7.8 && 0.2 &  6.9 &  6.0 &  5.4 \\
    000089 && 0.3 &  5.7 &  7.8 & 10.5 && 0.5 &  7.0 &  9.6 &  9.8 \\
    000406 && 0.6 &  8.3 & 19.4 & 32.7 && 0.9 & 10.8 & 22.6 & 26.0 \\
    000429 && 0.5 & 12.2 & 21.3 & 31.5 && 0.5 & 11.8 & 20.1 & 22.9 \\
    000488 && 0.7 & 12.1 & 19.4 & 29.1 && 0.7 & 12.0 & 22.3 & 23.5 \\
    000539 && 0.4 & 11.4 & 16.4 & 26.3 && 0.7 & 12.5 & 19.2 & 18.7 \\
    000541 && 0.8 & 14.6 & 18.7 & 26.8 && 0.7 & 17.8 & 22.7 & 19.6 \\
    000550 && 0.8 & 14.4 & 24.0 & 40.2 && 1.1 & 11.7 & 25.5 & 33.0 \\
    000581 && 0.8 & 16.1 & 17.0 & 28.4 && 0.9 & 16.9 & 19.3 & 19.6 \\
    000625 && 0.8 & 12.4 & 23.9 & 39.0 && 0.8 & 13.1 & 24.0 & 34.3 \\
    000709 && 0.5 &  9.0 & 18.0 & 30.6 && 0.4 &  7.3 & 19.2 & 21.7 \\
    000720 && 0.2 & 14.6 & 26.3 & 25.4 && 0.4 & 10.4 & 17.3 & 12.3 \\
    000778 && 0.4 & 10.8 & 18.6 & 29.6 && 0.5 & 12.9 & 20.6 & 20.1 \\
  \hline\hline
 \end{tabular}
\end{table}

It is shown that the cancellation ratio decreases with the submission aggressiveness. For the 23 stocks analyzed, we obtain the mean values $\overline{r}_1=0.5\%$, $\overline{r}_2=11.1\%$, $\overline{r}_3=18.2\%$ and $\overline{r}_4=27.7\%$ for buy orders, and the mean values $\overline{r}_1=0.6\%$, $\overline{r}_2=10.3\%$, $\overline{r}_3=19.2\%$ and $\overline{r}_4=21.3\%$ for sell orders. We find that a very small proportion of partially filled orders have been cancelled and more aggressive orders are less likely to be cancelled. These findings indicate that traders are more likely to cancel orders that have lower chance to be executed to reduce the NE risk.

\section{The profile of cancellation positions in the limit-order book}
\label{sec:PDF-LOB}

In this section, we study the profile of cancellation positions (probability distribution of cancellation position) in the LOB. Denote $x(t)$ as the price level where a cancellation occurred in the LOB at time $t$. Here we use the event time $t$, {\it{i.e.}} when a cancellation happens, it increases by one ($t \to t+1$). Suppose an order marked with gray color in figure~\ref{Fig:LOB} is cancelled at time $t$. The cancelled order is stored at the second price level in the buy LOB, {\it{i.e.}} $x(t)=2$. Generally, $x(t)=n$ indicates that a cancellation occurs at the $n$-th price level in the buy or sell LOB at time $t$.

The length of LOB, $L_b(t)$ or $L_s(t)$, defined as the number of price levels existing in the buy or sell LOB, plays an important role in the order cancellation process. For example, it is a completely different situation that a trader cancels an order at the second price level $x(t)=2$ when $L(t)=5$ from the situation that she cancels an order at the second price level when $L(t)=50$. Since it is trivial to study the position profile of the absolute values of $x(t)$, we will focus on the relative price levels $X(t)$ instead, which is defined as follows
\begin{equation}
X(t)=\frac{x(t)}{L_{b,s}(t)}~,
\label{Eq:X}
\end{equation}
where $X(t)$ varies in the range $\left(0,1\right]$. According to the definition, a small value of $X$ means that the cancelled order located in the front of LOB in reference to the best bid or ask price, while a large value of $X$ suggests that the cancellation happens far from the best price.

\begin{figure}[htb]
  \centering
  \includegraphics[width=6cm]{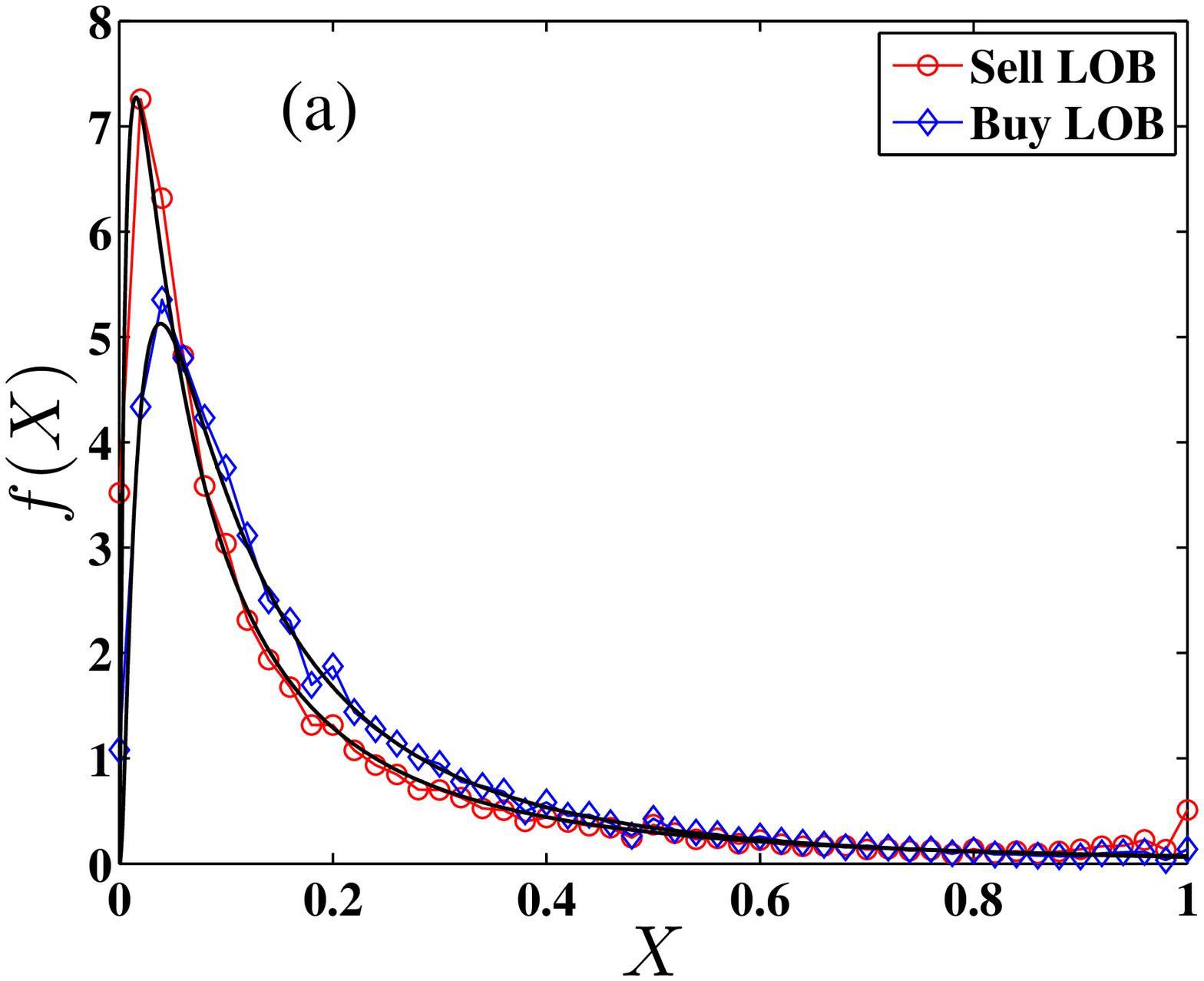}
  \includegraphics[width=6cm]{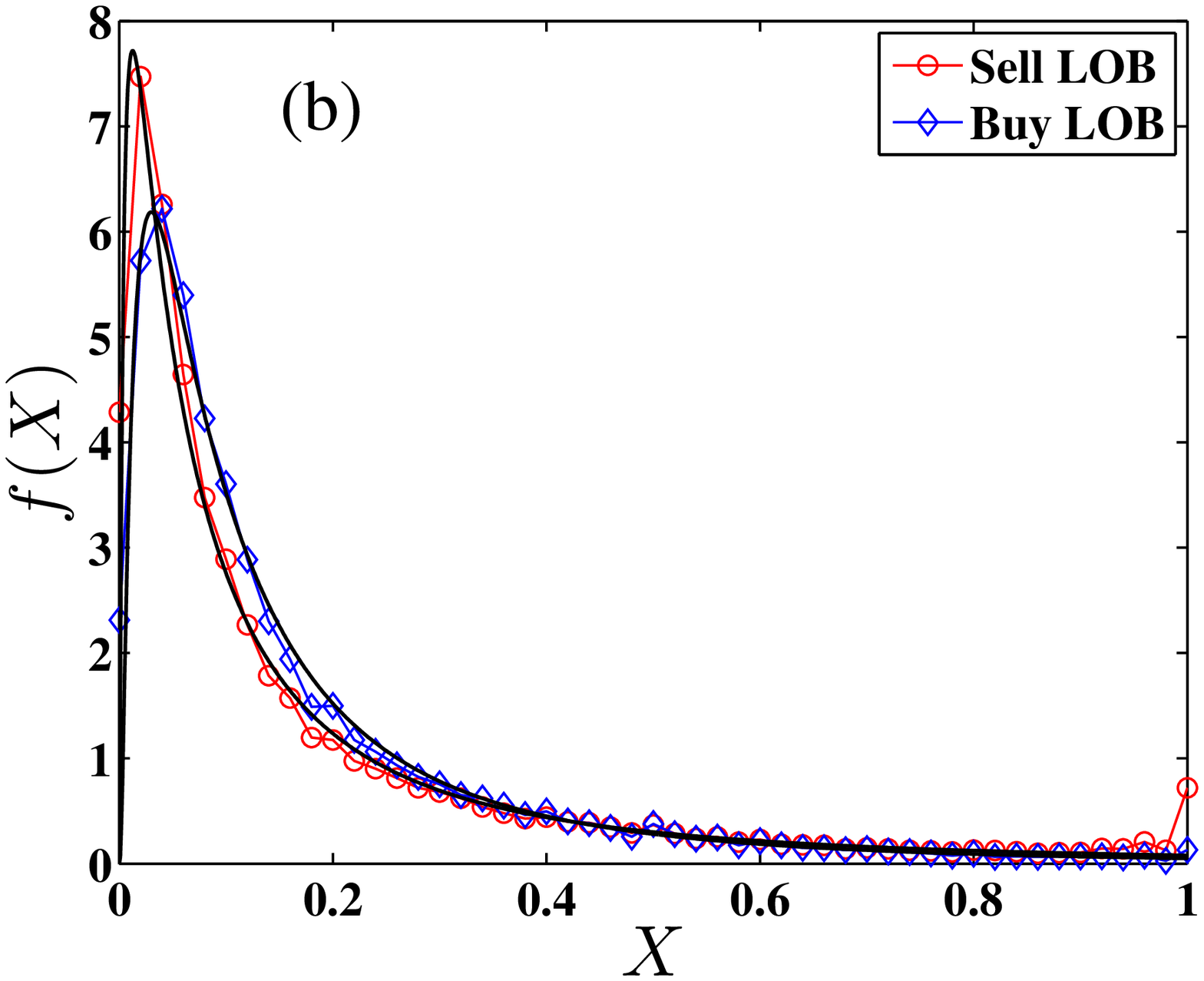}
  \includegraphics[width=6cm]{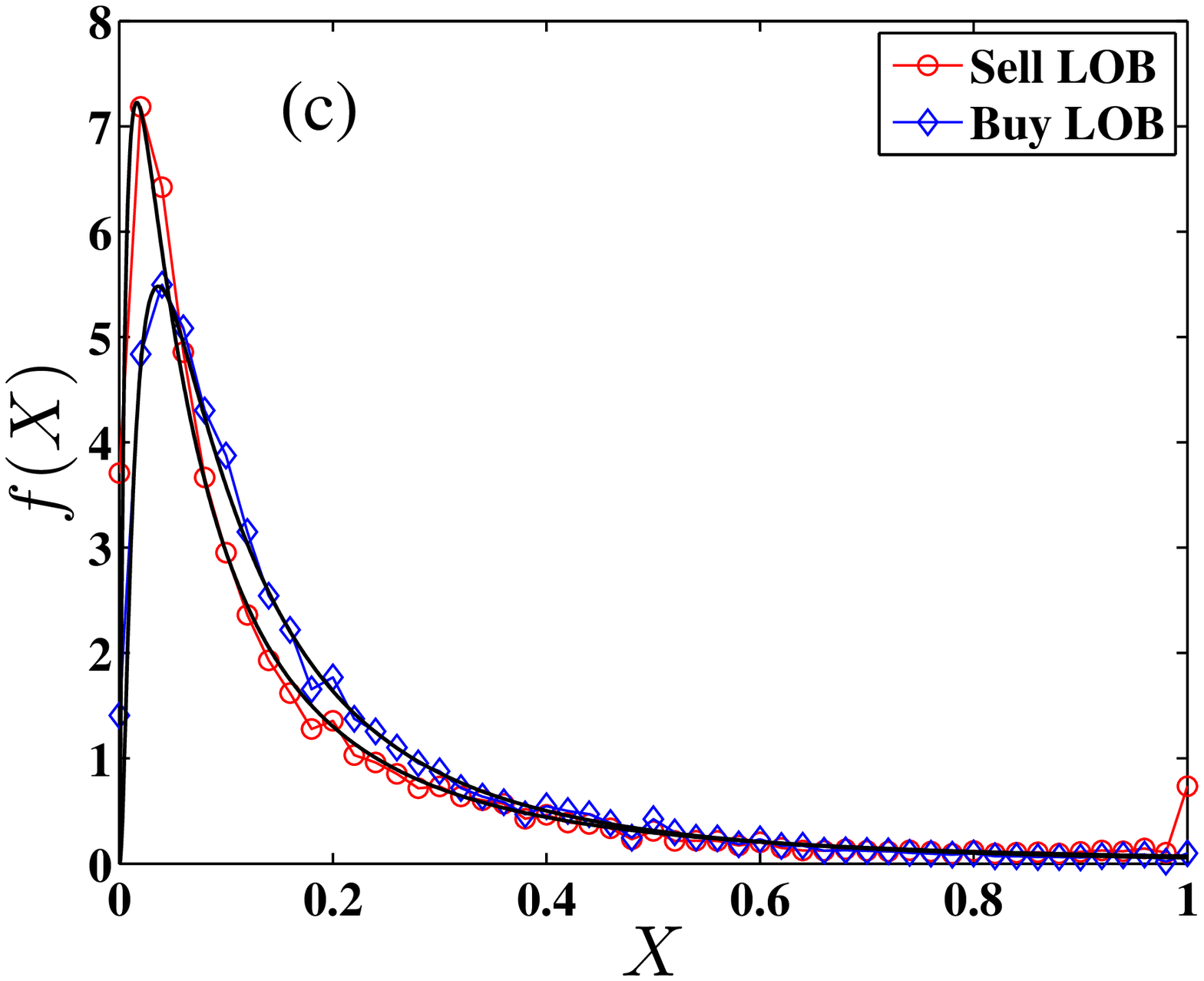}
  \includegraphics[width=6cm]{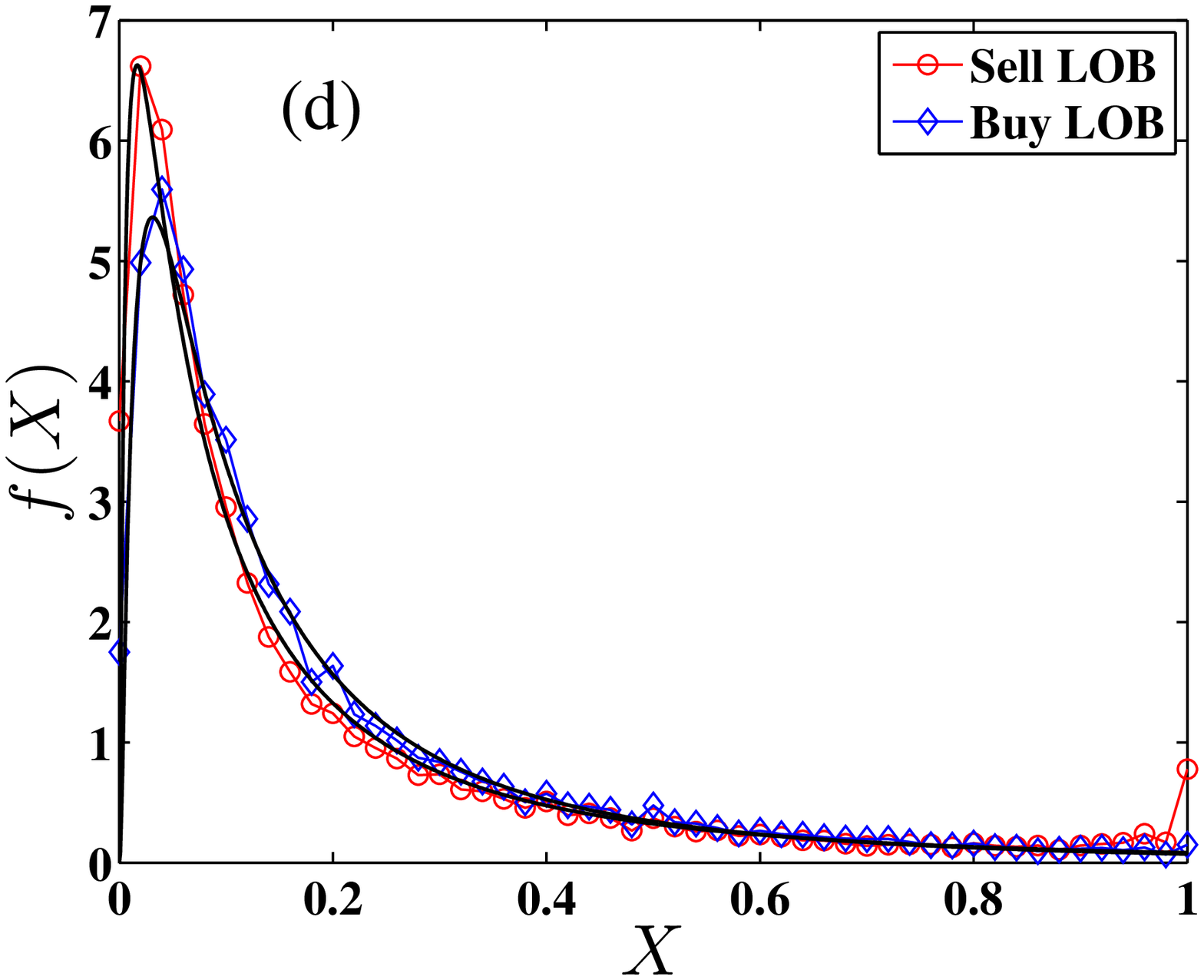}
  \caption{Probability density functions $f(X)$ of relative price levels of cancellations for both buy and sell orders of four randomly chosen stocks: (a) 000012, (b) 000021, (c) 000066 and (d) 0000625. The solid lines are fits to the log-normally distribution.}
  \label{Fig:ps_lob}
\end{figure}

We adopt the probability density function (PDF) to describe the position profile of order cancellations, and analyze the PDF $f(X)$ of relative price levels of cancellations for both buy and sell orders. In figure~\ref{Fig:ps_lob}, we illustrate the probability density functions $f(X)$ for four stocks randomly chosen from the 23 stocks analyzed.

As we know, the order stored in the LOB is executed only when all the orders ahead have been removed, so the orders at the end of the LOB are less to be executed and its cancellation probabilities increase as well. We expect that probability density function $f(X)$ should increase monotonically with respect to the price level $X$. However, as presented in figure~\ref{Fig:ps_lob}, the PDFs $f(X)$ first increase sharply, and then fall down rapidly for both buy and sell orders, which deviates from our expectation. The special shape of PDF may be cause by the reason that in the Chinese stock market many impatient traders place aggressive orders close to the same best price to increase the transaction probability, which is confirmed by the fact that the limit-order book shape has a maximum away from the same best price ($X=1$) but extremely close to the same best price \cite{Gu-Chen-Zhou-2008c-PA}. Once the impatient traders can not make an immediate transaction, they will cancel the placed orders and place a new order to catch the price in order to avoid NE risk, which results in a large cancellation probability in the front of the LOB. In addition, the PDFs $f(X)$ are not identical between buy orders and sell orders. For sell orders, they have higher cancellation probability in the front and at the end of the sell LOB, however buy orders have higher cancellation probability in the middle of the buy LOB. The sharply increasing value of $f(X)$ located at $X=1$ may be caused by the 10\% price limit trading rules in the Chinese stock market.

In figure~\ref{Fig:ps_lob}, we observe that the PDF $f(X)$ of relative price levels follows a log-normal distribution. Since the relative price $X$ varies in the range $(0, 1]$, we normalize the log-normal distribution to fit the PDF, that is,
\begin{equation}
f(X)=\frac{1}{z}\frac{1}{\sqrt{2\pi}\sigma{X}}\exp\left[-\frac{(\ln{X}-\mu)^2}{2\sigma^2}\right]~,
\label{Eq:logn}
\end{equation}
where $\mu$ is the location parameter and $\sigma$ is the scaling parameter, and $z$ is a normalized constant ($z=\int_0^1(\frac{1}{\sqrt{2\pi}\sigma{x}}\exp\left[-\frac{(\ln{x}-\mu)^2}{2\sigma^2}\right]{\rm{d}}x$). We analyze the remaining stocks as well, and find that their PDFs are all log-normally distributed. Using the least squares fitting method, we calculate the parameters $\mu$ and $\sigma$ for both buy and sell orders of 23 stocks, which are listed in table~\ref{Tb:Lognormal}. In order to compare the performance of log-normal distribution fit, we introduce a new statistical variable $rms$ (defined as the root mean square (r.m.s) of the difference between the best fit and the empirical data) which is also depicted in table~\ref{Tb:Lognormal}. We also apply gamma distribution to fit the data and find that log-normal distribution gives better results.

\begin{table}[htp]
\centering
\caption{Characteristic parameters of PDFs $f(X)$ for both buy and sell orders of 23 stocks. $\mu$ and $\sigma$ are the parameters of log-normal distribution obtained by the least squares fitting method. $rms$ is the r.m.s. of the difference between the best fit and the empirical data. $Prob$ is the p-value calculated from Monte Carlo method with 1000 repeats.}
\medskip
\label{Tb:Lognormal}
\centering
\begin{tabular}{ccccccccccc}
 \hline \hline
 \multirow{3}*[2mm]{Stock} & \multicolumn{4}{@{\extracolsep\fill}c}{Buy orders} &&& \multicolumn{4}{@{\extracolsep\fill}c}{Sell orders}\\
 \cline{2-5} \cline{8-11}
 & \multicolumn{1}{@{\extracolsep\fill}c}{$\mu$} & \multicolumn{1}{@{\extracolsep\fill}c}{$\sigma$} & \multicolumn{1}{@{\extracolsep\fill}c}{$rms$} & \multicolumn{1}{@{\extracolsep\fill}c}{$Prob$} &&& \multicolumn{1}{@{\extracolsep\fill}c}{$\mu$} & \multicolumn{1}{@{\extracolsep\fill}c}{$\sigma$} & \multicolumn{1}{@{\extracolsep\fill}c}{$rms$} & \multicolumn{1}{@{\extracolsep\fill}c}{$Prob$} \\
  \hline
    000001 & -2.36 & 1.13 & 0.13 & 0.00 &&& -2.49 & 1.52 & 0.09 & 0.00 \\
    000002 & -2.17 & 1.10 & 0.33 & 0.89 &&& -2.37 & 1.22 & 0.11 & 0.00 \\
    000009 & -2.26 & 1.01 & 0.21 & 0.28 &&& -2.27 & 1.26 & 0.16 & 0.15 \\
    000012 & -2.04 & 1.10 & 0.07 & 0.00 &&& -2.28 & 1.37 & 0.10 & 0.00 \\
    000016 & -2.00 & 1.07 & 0.16 & 0.01 &&& -2.26 & 1.27 & 0.11 & 0.00 \\
    000021 & -2.22 & 1.13 & 0.09 & 0.00 &&& -2.29 & 1.45 & 0.11 & 0.00 \\
    000024 & -1.93 & 1.13 & 0.10 & 0.00 &&& -2.25 & 1.30 & 0.10 & 0.00 \\
    000027 & -2.14 & 1.04 & 0.10 & 0.00 &&& -2.26 & 1.23 & 0.13 & 0.00 \\
    000063 & -2.02 & 1.22 & 0.08 & 0.00 &&& -2.26 & 1.50 & 0.13 & 0.00 \\
    000066 & -2.11 & 1.10 & 0.07 & 0.00 &&& -2.29 & 1.35 & 0.11 & 0.00 \\
    000088 & -1.58 & 1.21 & 0.25 & 0.68 &&& -2.06 & 1.43 & 0.08 & 0.00 \\
    000089 & -2.01 & 1.14 & 0.17 & 0.06 &&& -2.32 & 1.23 & 0.12 & 0.00 \\
    000406 & -2.25 & 1.03 & 0.09 & 0.00 &&& -2.40 & 1.26 & 0.15 & 0.00 \\
    000429 & -1.99 & 0.98 & 0.18 & 0.02 &&& -2.33 & 1.18 & 0.09 & 0.00 \\
    000488 & -1.88 & 1.26 & 0.11 & 0.00 &&& -2.12 & 1.34 & 0.11 & 0.00 \\
    000539 & -1.80 & 1.14 & 0.21 & 0.37 &&& -2.09 & 1.37 & 0.09 & 0.00 \\
    000541 & -1.80 & 1.23 & 0.18 & 0.20 &&& -2.40 & 1.39 & 0.15 & 0.04 \\
    000550 & -2.01 & 1.13 & 0.08 & 0.00 &&& -2.21 & 1.34 & 0.12 & 0.00 \\
    000581 & -1.75 & 1.13 & 0.17 & 0.08 &&& -2.24 & 1.35 & 0.10 & 0.00 \\
    000625 & -2.07 & 1.18 & 0.10 & 0.00 &&& -2.18 & 1.38 & 0.13 & 0.04 \\
    000709 & -2.13 & 1.00 & 0.12 & 0.00 &&& -2.53 & 1.11 & 0.16 & 0.00 \\
    000720 & -1.77 & 1.31 & 0.16 & 0.09 &&& -1.79 & 1.56 & 0.16 & 0.11 \\
    000778 & -2.02 & 1.11 & 0.12 & 0.00 &&& -2.33 & 1.38 & 0.12 & 0.01 \\
  \hline\hline
 \end{tabular}
\end{table}

According to table \ref{Tb:Lognormal}, the value of $\mu$ obtained form cancelled sell orders is less than cancelled buy orders for each stock, while the value of $\sigma$ of cancelled sell orders is greater than cancelled buy orders. We also calculate the mean values of $\mu$ and $\sigma$ and find that $\overline{\mu}=-2.01\pm0.18$ and $\overline{\sigma}=1.13\pm0.09$ for buy orders and $\overline{\mu}=-2.26\pm0.15$ and $\overline{\sigma}=1.34\pm0.11$ for sell orders. It means that there exists a higher peak in the PDF for cancelled sell orders and the peak for cancelled sell orders is closer to the same best price, which implies that sellers are more eager to make an immediate transaction. This market behavior is consistent with the situation that it was a bear stock market during the year 2003 and market participants were more willing to sell their orders.

Since the majority of PDFs $f(X)$ of the 23 stocks are log-normally distributed as shown in table \ref{Tb:Lognormal}, we aggregate all the data together and treat them as an ensemble to obtain a better statistic. As expected, the PDFs of ensemble relative price levels obey the log-normal distribution for both buy and sell orders. Using the same method, we obtain the parameters $\mu=-2.14$ and $\sigma=1.11$ with $rms=0.07$ for ensemble buy orders, and $\mu=-2.29$ and $\sigma=1.35$ with $rms=0.11$ for ensemble sell orders, which are all close to their mean values mentioned above.

Except for the factor of price levels in the LOB, the number of orders stored in the LOB also affects traders' cancellation strategy, especially for the number of orders stored at the price level where the cancellation takes place. Denote $n(x,t)$ as the number of orders stored at price level $x$ at time $t$. We conjecture that the cancellation probability of an order at price level $x(t)$ is positively correlated with $n(x,t)$ because a long queue of orders at the same price level of an order means a low chance for the order to be executed. Our purpose is to analyze the relation between the cancellation probability and the cancellation position in the LOB, and it is necessary to remove the number effect at each price level. We define the normalized relative price levels $\widehat{X}(t)$ by
\begin{equation}
\widehat{X}(t)=X(t)\left/\frac{n(x,t)}{\sum_{x=1}^{L_{b,s}(t)}n(x,t)}\right.~,
\label{Eq:wX}
\end{equation}
where $L_{s,b}(t)$ is the length of buy or sell LOB and $\sum_{x=1}^{L_{b,s}(t)}{n(x,t)}$ is the total number of orders stored in the buy or sell LOB at time $t$. This normalization aims at flatten the order book shape, which usually has a hump for stocks in the Chinese market \cite{Gu-Chen-Zhou-2008c-PA} and in other markets as well \cite{Bouchaud-Mezard-Potters-2002-QF,Potters-Bouchaud-2003-PA,Challet-Stinchcombe-2001-PA,Maslov-Mills-2001-PA,Weber-Rosenow-2005-QF,Eisler-Kertesz-Lillo-2007-PSPIE}. In this way, the numbers of orders at different price levels can be viewed to be identical.

\begin{figure}[htb]
  \centering
  \includegraphics[width=6cm]{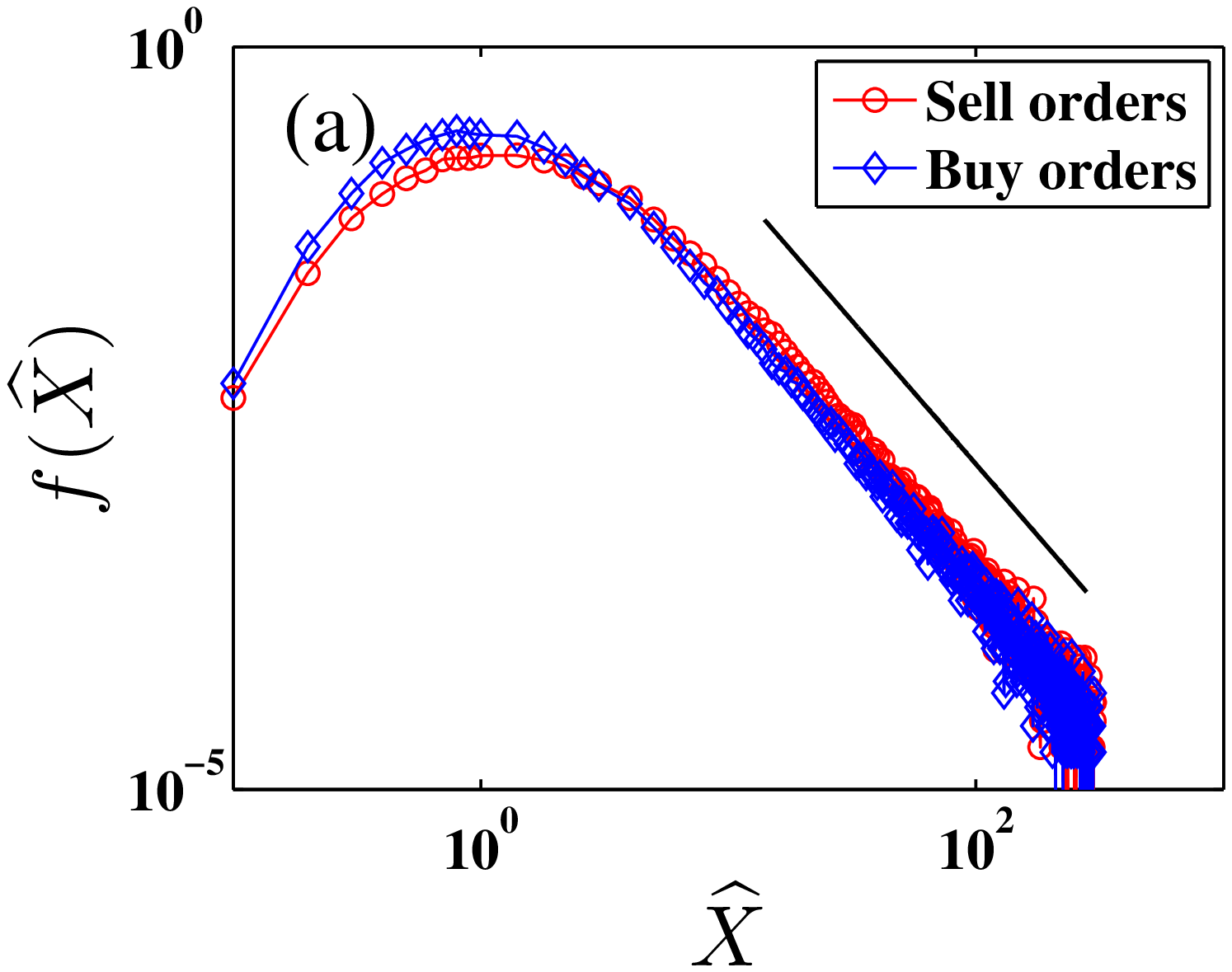}
  \includegraphics[width=6cm]{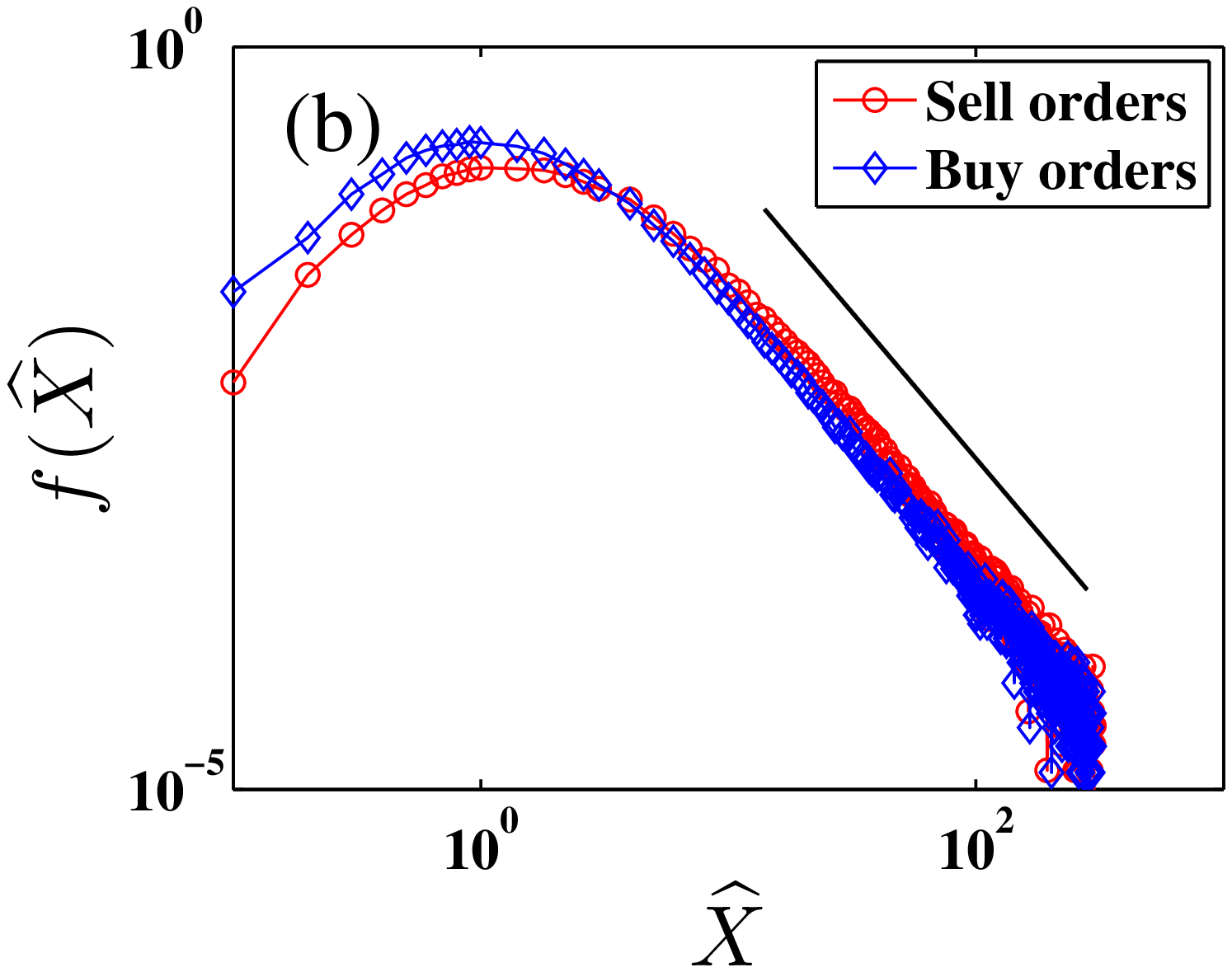}
  \includegraphics[width=6cm]{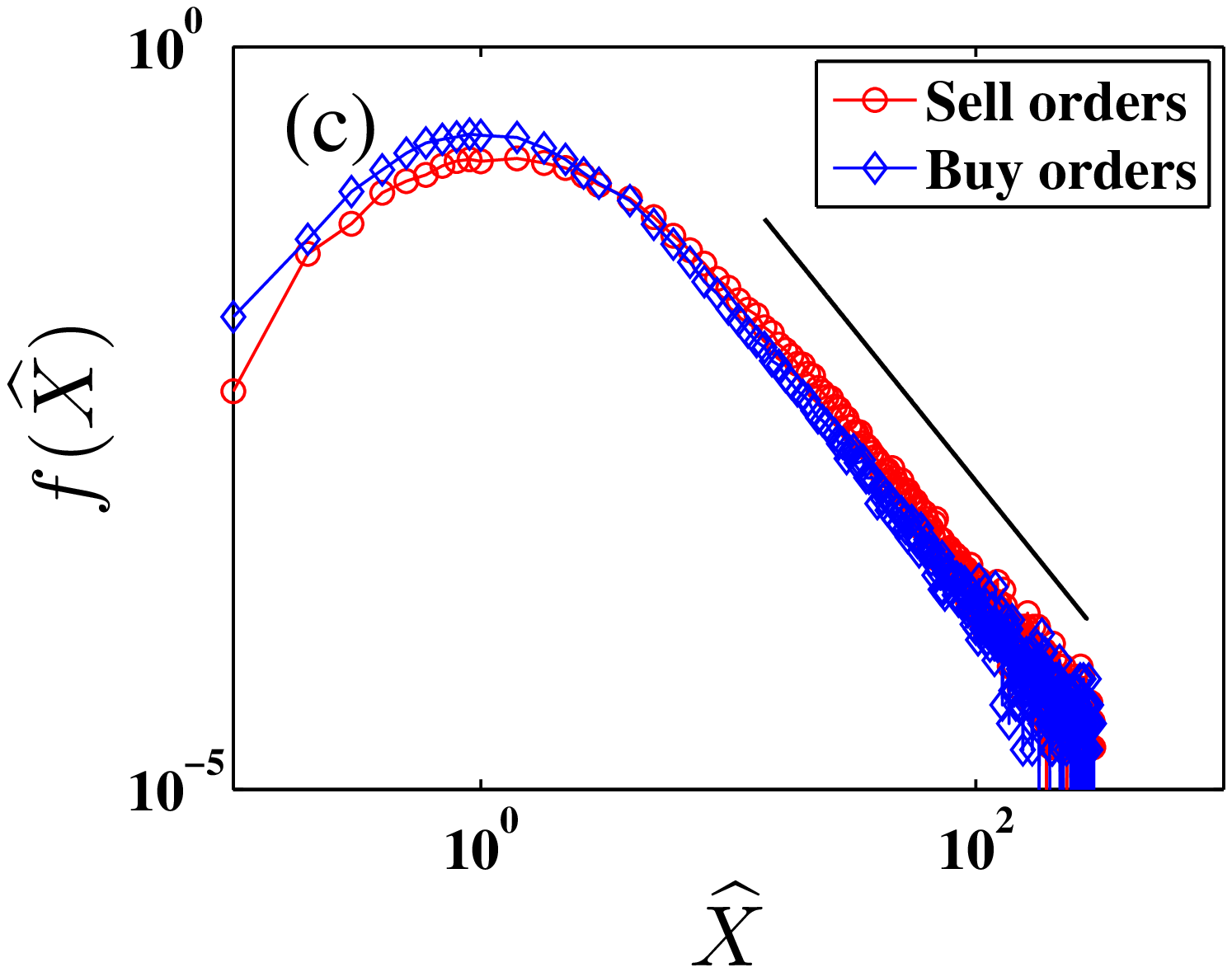}
  \includegraphics[width=6cm]{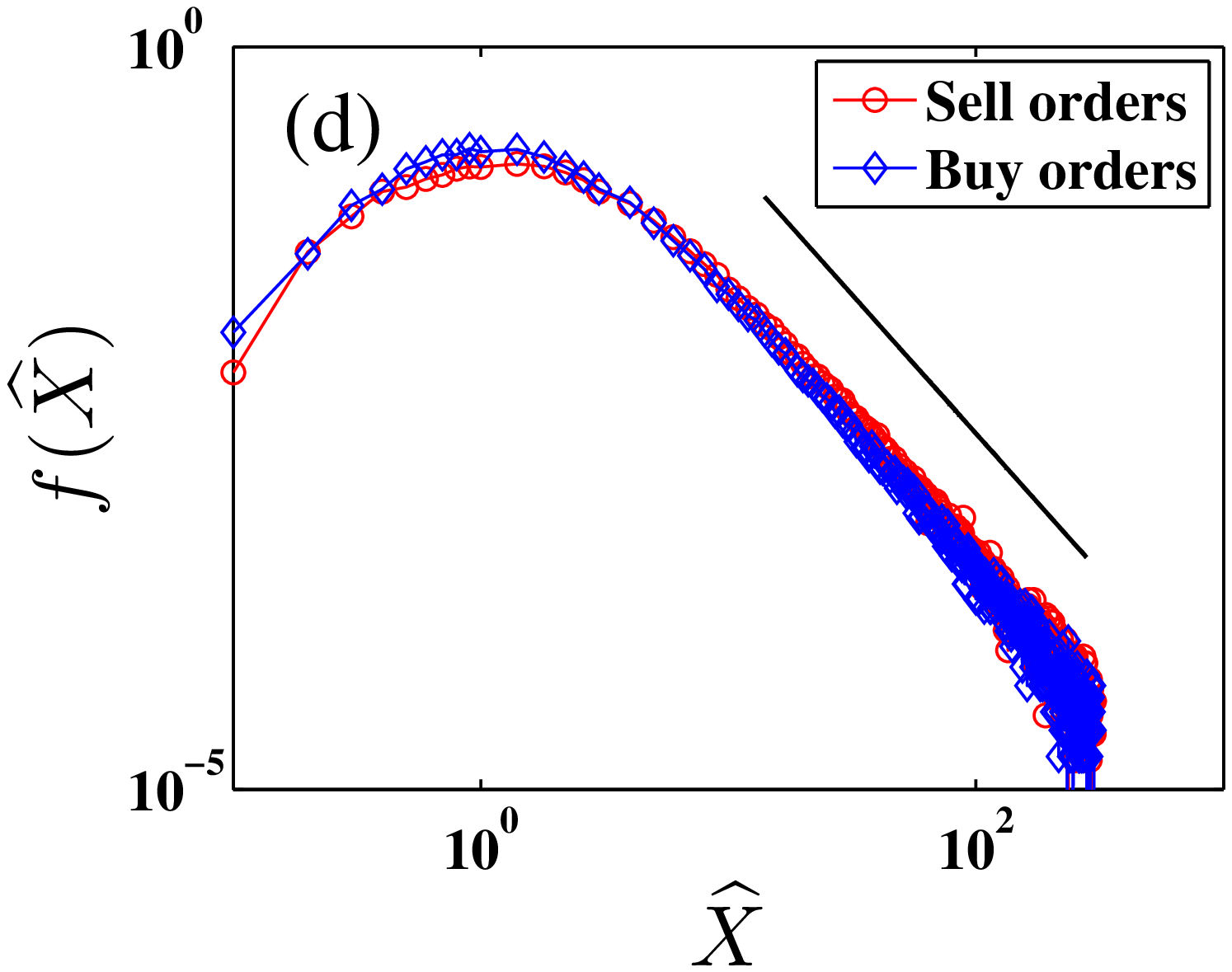}
  \caption{Empirical probability density functions $f(\widehat{X})$ at log-log scales for both cancelled buy and sell orders of four stocks: (a) 000012, (b) 000021, (c) 000066 and (d) 0000625. The solid lines are the fitted functions using the maximum likelihood estimation method.}
  \label{Fig:ps_lob_nm}
\end{figure}

In figure~\ref{Fig:ps_lob_nm}, we present the probability density functions $f(\widehat{X})$ of the normalized relative price levels $\widehat{X}$ for both cancelled buy and sell orders of the same four stocks as in figure \ref{Fig:ps_lob}. We find that the PDFs first increase with the normalized relative price level $\widehat{X}$, and then decrease rapidly for both buy and sell orders. Each PDF has its maximum value away from the same best price, which is in line with the results presented in figure~\ref{Fig:ps_lob}. In addition, the PDF follows a power-law decay when $\widehat{X}>\widehat{X}_{\min}$,
\begin{equation}
f(\widehat{X}>\widehat{X}_{\min})\sim{\widehat{X}}^{-\alpha},
\label{Eq:fwX}
\end{equation}
where $\alpha$ is the tail exponent of the power law and $\widehat{X}_{\min}$ is the lower threshold of the scaling range of the power law decay.

There are many methods to determine the power-law tail exponent $\alpha$. Clauset, Shalizi and Newman proposed an efficient quantitative method to estimate $\widehat{X}_{\min}$ and $\alpha$ based on the Kolmogorov-Smirnov test. The Kolmogorov-Smirnov statistic ($KS$) is defined as
\begin{equation}
  KS=\max_{\widehat{X}>\widehat{X}_{\min}}(|P-F_{\rm{PL}}|),
  \label{Eq:ks}
\end{equation}
where $P$ is the cumulative distribution of normalize relative price levels $\widehat{X}$ and $F_{\rm{PL}}$ is the cumulative distribution of the best power-law fit \cite{Clauset-Shalizi-Newman-2009-SIAMR}. One can first determine the threshold $\widehat{X}_{\min}$ by minimizing the $KS$ statistic, and then estimate the power-law tail exponent $\alpha$ with the data in the range $\widehat{X}>\widehat{X}_{\min}$ using the maximum likelihood estimation (MLE) method,
\begin{equation}
  \alpha=1+m\left[\sum_{t=1}^{m}{\ln\frac{\widehat{X}(t)}{\widehat{X}_{\min}}}\right]^{-1},
  \label{Eq:alpha}
\end{equation}
where $m$ is the number of the data points in the range $\widehat{X}>\widehat{X}_{\min}$. The standard error $\hat{\sigma}$ is derived from the width of the likelihood maximum, which can be expressed as follows
\begin{equation}
\hat{\sigma}=\frac{\alpha-1}{\sqrt{m}}.
\label{Eq:err}
\end{equation}

We analyze the PDFs of normalized price levels $\widehat{X}$ for other stocks and find that they all exhibit power-law right tails. Applying the method mentioned above, we determine the characteristic parameters $\alpha$, $\widehat{X}_{\min}$ and $\hat{\sigma}$, which are illustrated in table~\ref{Tb:LnPl}. We find that the power-law exponent $\alpha$ varies in the range $[1.80,~2.51]$ for buy cancellations and $[1.74,~2.42]$ for sell cancellations. The exponents $\alpha$ are all close to 2, with the mean value $\overline\alpha=2.06\pm0.16$ for cancelled buy orders and $\overline\alpha=2.12\pm0.17$ for cancelled sell orders. Moreover, for 8 stocks, the tail exponents of the cancelled buy orders are greater than the cancelled sell orders.

\begin{table}[htp]
\centering
\caption{Characteristic parameters of PDFs of normalized relative price levels $\widehat{X}$ for both buy and sell cancellations of the 23 stocks. $\alpha$, $\widehat{X}_{\min}$ and $\hat{\sigma}$ are the parameters of power-law tail distribution based on the KS-test and likelihood maximum estimation method.}
\medskip
\label{Tb:LnPl}
\centering
\begin{tabular}{crrrrrrrr}
 \hline \hline
 \multirow{3}*[2mm]{Stock} & \multicolumn{3}{@{\extracolsep\fill}c}{Buy orders} &&& \multicolumn{3}{@{\extracolsep\fill}c}{Sell orders}\\
 \cline{2-4} \cline{7-9}
 & \multicolumn{1}{@{\extracolsep\fill}c}{$\alpha$} & \multicolumn{1}{@{\extracolsep\fill}c}{$\hat{\sigma}$} & \multicolumn{1}{@{\extracolsep\fill}c}{$\widehat{X}_{\min}$} &&& \multicolumn{1}{@{\extracolsep\fill}c}{$\alpha$} & \multicolumn{1}{@{\extracolsep\fill}c}{$\hat{\sigma}$} & \multicolumn{1}{@{\extracolsep\fill}c}{$\widehat{X}_{\min}$} \\
  \hline
    000001 & 1.82 & 0.001 &  4.85 &&& 2.09 & 0.002 & 10.61 \\
    000002 & 1.96 & 0.005 &  4.07 &&& 2.16 & 0.006 &  8.54 \\
    000009 & 2.01 & 0.002 &  4.06 &&& 2.35 & 0.003 & 10.41 \\
    000012 & 1.96 & 0.003 &  5.63 &&& 2.07 & 0.003 &  9.18 \\
    000016 & 2.07 & 0.004 &  5.04 &&& 2.19 & 0.005 & 11.10 \\
    000021 & 2.05 & 0.003 & 11.23 &&& 2.19 & 0.003 &  9.75 \\
    000024 & 2.10 & 0.005 &  7.75 &&& 1.90 & 0.004 &  3.94 \\
    000027 & 2.08 & 0.006 &  7.73 &&& 2.06 & 0.005 & 10.88 \\
    000063 & 2.18 & 0.007 & 18.49 &&& 2.26 & 0.007 & 19.34 \\
    000066 & 2.10 & 0.003 &  9.79 &&& 2.22 & 0.004 & 18.98 \\
    000088 & 2.51 & 0.026 & 19.39 &&& 2.42 & 0.028 & 22.00 \\
    000089 & 2.11 & 0.008 &  6.36 &&& 1.93 & 0.006 &  3.68 \\
    000406 & 2.01 & 0.003 &  4.89 &&& 2.13 & 0.004 & 19.40 \\
    000429 & 2.10 & 0.006 &  3.30 &&& 2.19 & 0.006 & 15.17 \\
    000488 & 2.45 & 0.012 & 15.53 &&& 1.98 & 0.005 & 17.29 \\
    000539 & 2.04 & 0.006 &  5.67 &&& 2.23 & 0.008 & 27.91 \\
    000541 & 1.95 & 0.007 &  3.79 &&& 1.79 & 0.006 &  2.76 \\
    000550 & 1.99 & 0.003 &  6.24 &&& 2.19 & 0.003 & 13.87 \\
    000581 & 2.12 & 0.007 &  7.86 &&& 2.01 & 0.006 & 11.04 \\
    000625 & 2.00 & 0.003 &  9.31 &&& 2.18 & 0.003 & 12.15 \\
    000709 & 2.02 & 0.004 &  3.26 &&& 2.23 & 0.005 & 12.01 \\
    000720 & 1.80 & 0.006 &  3.35 &&& 1.74 & 0.006 &  3.83 \\
    000778 & 2.07 & 0.005 &  7.55 &&& 2.19 & 0.005 & 18.74 \\
  \hline\hline
 \end{tabular}
\end{table}

Because the PDFs of normalized relative price levels $\widehat{X}$ for the 23 stocks are similar to one another, we treat the 23 stocks as an ensemble and determine the PDF with the 23-stock aggregating data. As expected, the ensemble PDF is similar to the individual stock, with a power-law relaxation in the right tail. Using the KS-test and maximum likelihood estimation method, we obtain $\alpha=2.02$ with $\widehat{X}_{\min}=7.34$ for cancelled buy orders and $\alpha=2.15$ with $\widehat{X}_{\min}=10.59$ for cancelled sell orders, which are close to the mean values ($\overline\alpha=2.06$ for buy cancellations and $\overline\alpha=2.12$ for sell cancellations), respectively.

\section{The profile of cancellation positions at the price level}
\label{sec:PDF-PL}

In this section, we study the profile of cancellation positions at a certain price level, and apply the probability density function (PDF) to analyze it as well. Denote $y(x,t)$ as the position of a cancelled order in the queue of the $x$-th price level at time $t$. For example, the cancelled order marked with gray color in figure~\ref{Fig:LOB} is located at the second position at the second price level ($x=2$) in the buy LOB at time $t$, {\it{i.e.}} $y(2,t)=2$. Taking into account the impact of the number of orders stored at the $x$-th price level, we define the relative position variable $Y(x,t)$ instead of $y(x,t)$,
\begin{equation}
  Y(x,t)=\frac{y(x,t)}{n_{b,s}(x,t)}~,
  \label{Eq:Y}
\end{equation}
where $Y(x,t)$ varies in the range $(0,1]$, and $n_{b,s}(x,t)$ is the total number of orders stored at the $x$-th price level at time $t$ in the buy or sell LOB.

\begin{figure}[htb]
\centering
\includegraphics[width=6cm]{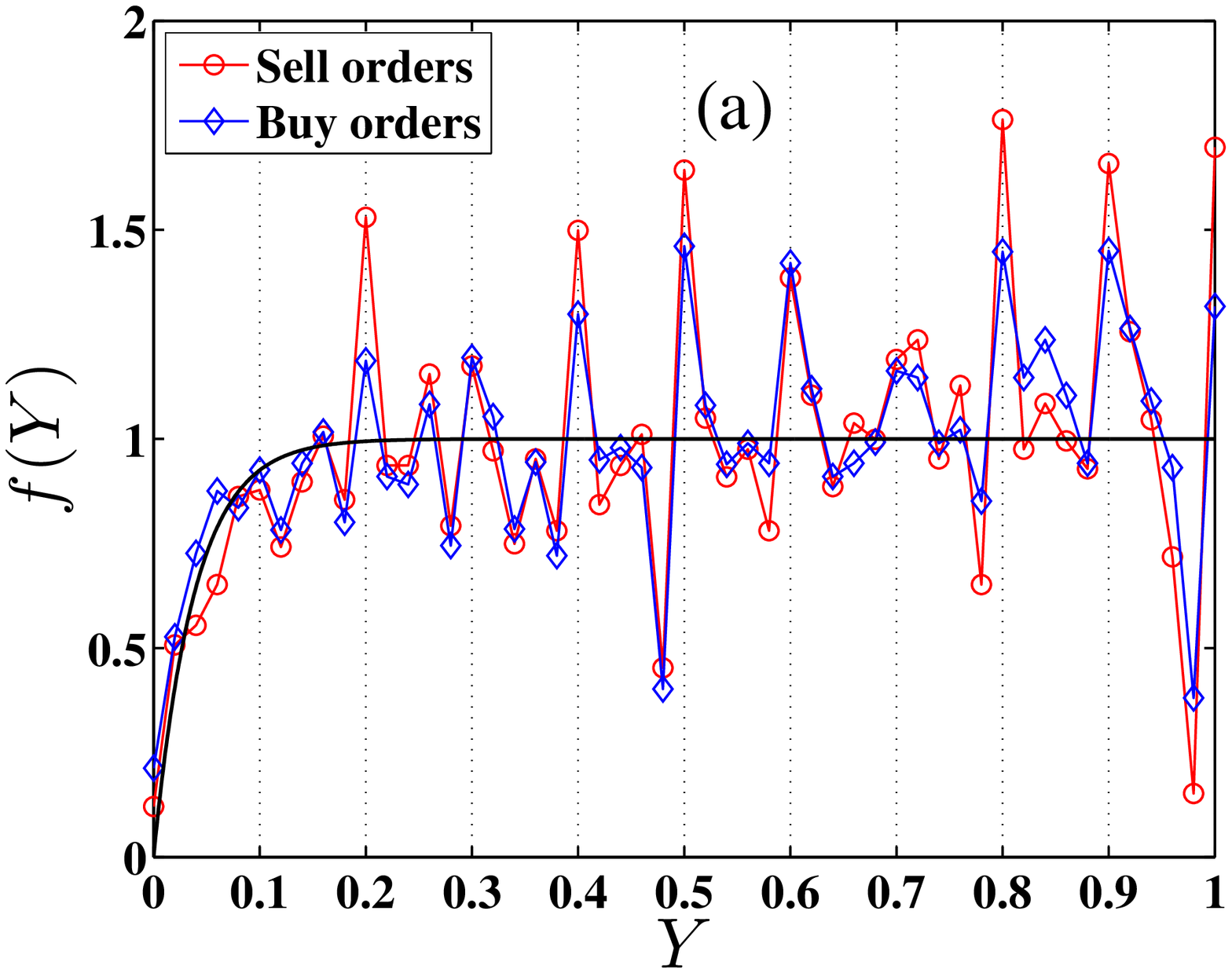}
\includegraphics[width=6cm]{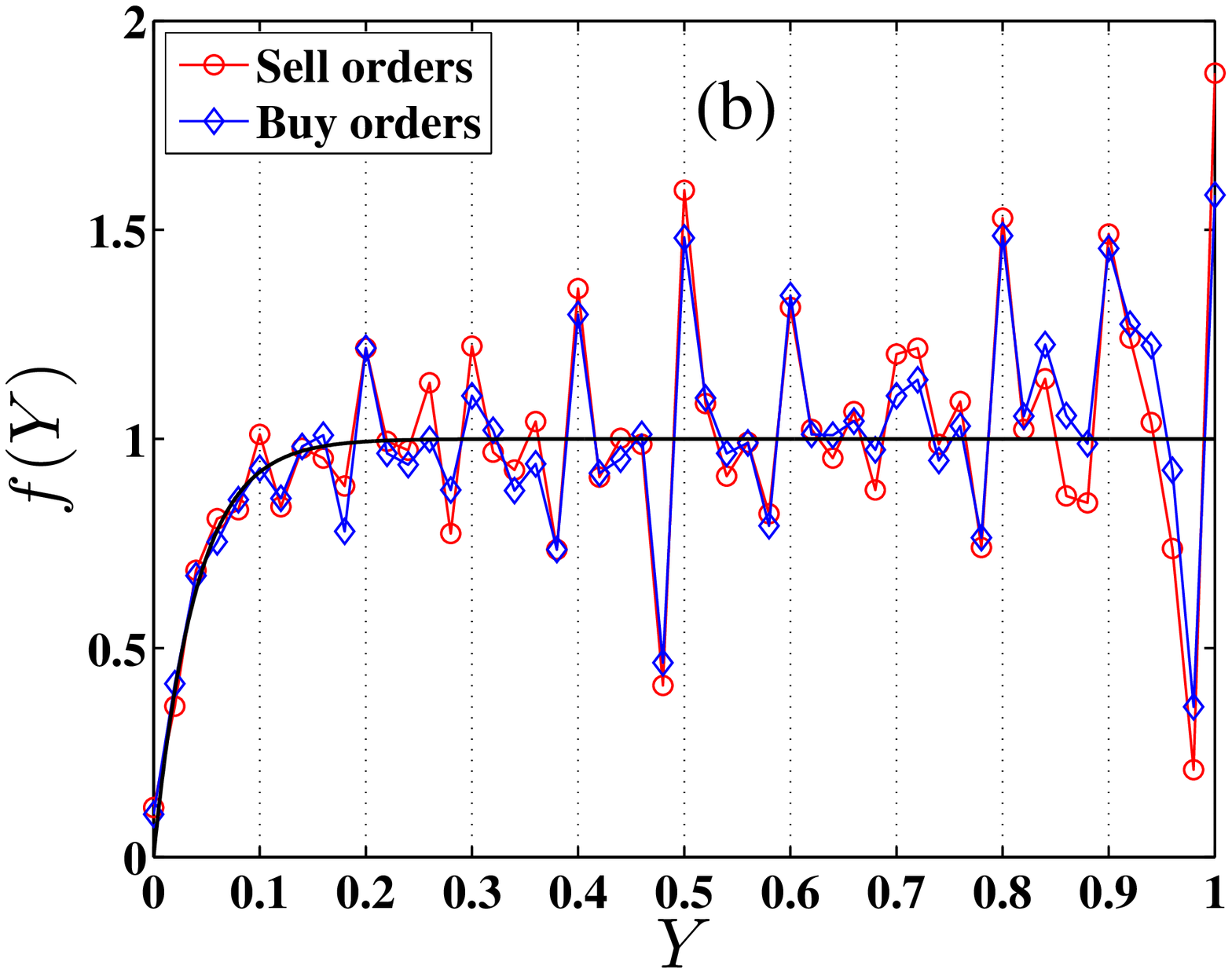}
\includegraphics[width=6cm]{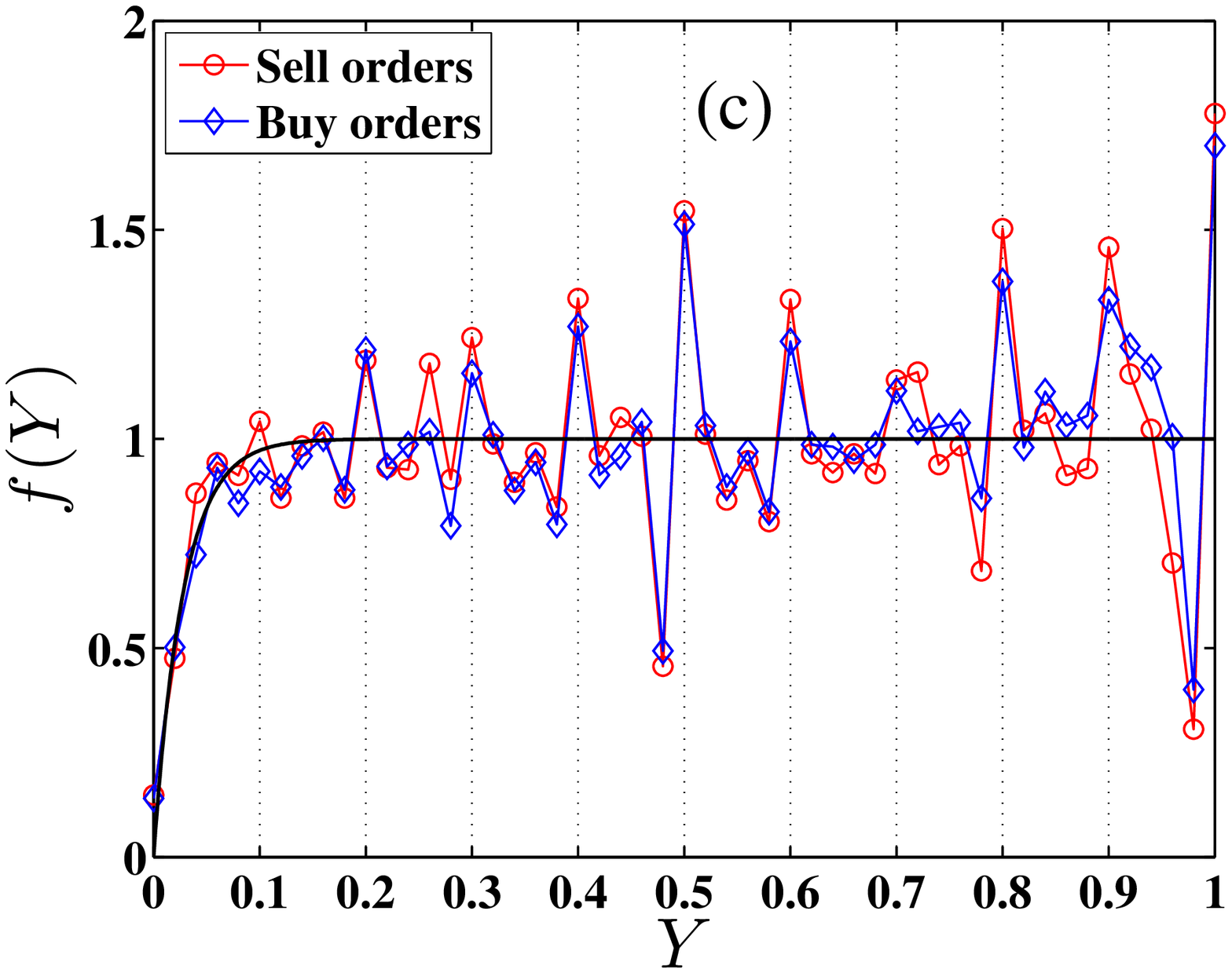}
\includegraphics[width=6cm]{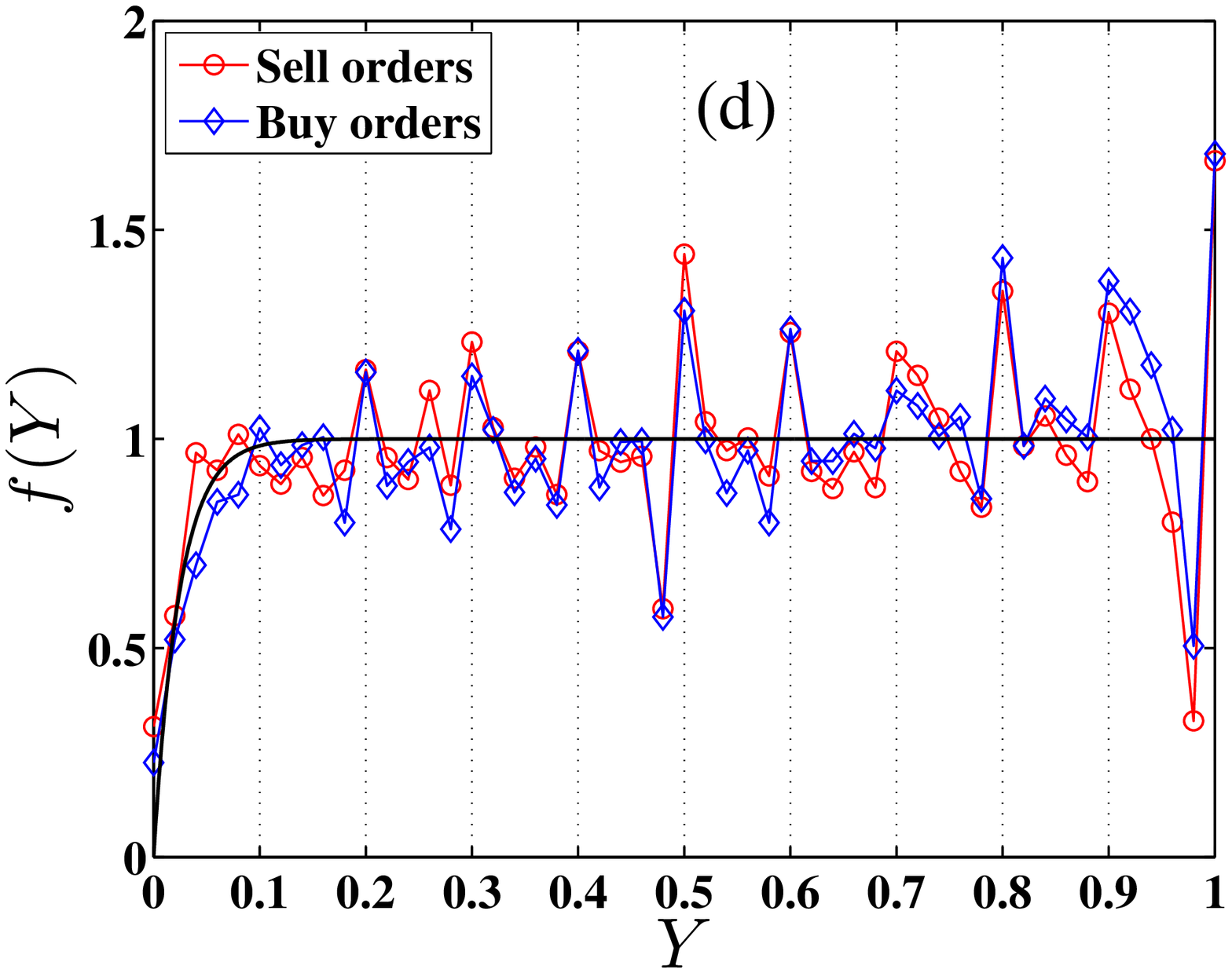}
\caption{Probability density functions $f(Y)$ of relative positions $Y(x,t)$ at the first price level ($x=1$) (a), the second price level ($x=2$) (b), the third price level ($x=3$) (c), and the fourth price level ($x=4$) (d) for both buy and sell orders of stock 000001. The solid curves are the best fitting function illustrated in Eq.~(\ref{Eq:fYz}).}
\label{Fig:ps_pl_pl}
\end{figure}

Now we analyze the probability density functions $f(Y)$ of relative positions $Y(x,t)$ at the $x$-th price level for both buy and sell cancellations. The PDFs $f(Y)$ of the first four price levels ($x=1,~2,~3,~4$) of stock 000001 are illustrated in figure~\ref{Fig:ps_pl_pl}, and we find the PDFs have similar shapes. Furthermore, the function $f(Y)$ is close to zero when the relative position $Y$ approaches to zero. As $Y$ increases, $f(Y)$ first increases rapidly in the range $Y\leq0.1$, then it fluctuates around a constat level until the end of the LOB. As we can see in the figures, the probability of order cancellation is small in the front of the queue at the first four price levels, which means that orders earlier submitted are less to be cancelled. It may be caused by the reason that in the Chinese stock market, most of the traders are patient and FO risk is not the main reason for cancellation.

In order to approximate the PDF of relative positions, we empirically apply an exponential function, which reads,
\begin{equation}
f(Y)=\frac{1}{z}(1-e^{{\beta}Y})~,
\label{Eq:fYz}
\end{equation}
where $\beta$ is request exponent and $z$ is a normalized constant, that is, $z=\int_0^1(1-e^{{\beta}x}){\rm{d}}x$. Using the least squares fitting method, we estimate the exponents $\beta$ of relative positions $Y(x,t)$($x=1,~2,~3,~4$) for both buy and sell orders of stock 000001, which are presented in table \ref{Tb:Lognormal-pl}.

\begin{table}[htp]
\centering
\caption{Characteristic parameters of PDFs of the first four price levels for both buy and sell orders of stock 000001. $\beta$ is the exponent of the fit function expressed in Eq.~(\ref{Eq:fYz}) and $rms$ is the r.m.s of the difference between the best fit and the empirical data mentioned in table \ref{Tb:Lognormal}.}
\medskip
\label{Tb:Lognormal-pl}
\centering
\begin{tabular}{crrrrc|ccrrrr}
 \hline \hline
 \multirow{3}*[2mm]{Price level} & \multicolumn{2}{c}{Buy orders} & \multicolumn{2}{c}{Sell orders} && \multirow{3}*[2mm]{Price level} & \multicolumn{2}{c}{Buy orders} & \multicolumn{2}{c}{Sell orders} \\
 \cline{2-5}\cline{8-11}
 & \multicolumn{1}{c}{$\beta$} & \multicolumn{1}{c}{$rms$} & \multicolumn{1}{c}{$\beta$} & \multicolumn{1}{c}{$rms$} &&& \multicolumn{1}{c}{$\beta$} & \multicolumn{1}{c}{$rms$} & \multicolumn{1}{c}{$\beta$} & \multicolumn{1}{c}{$rms$} \\
  \hline
    $x=1$ & -30.34 & 0.22 & -21.51 & 0.30 && $x=3$ & -22.25 & 0.21 & -29.54 & 0.24 \\
    $x=2$ & -24.89 & 0.22 & -25.78 & 0.26 && $x=4$ & -25.82 & 0.20 & -20.35 & 0.20 \\
  \hline\hline
 \end{tabular}
\end{table}

\begin{figure}[htb]
\centering
\includegraphics[width=6cm]{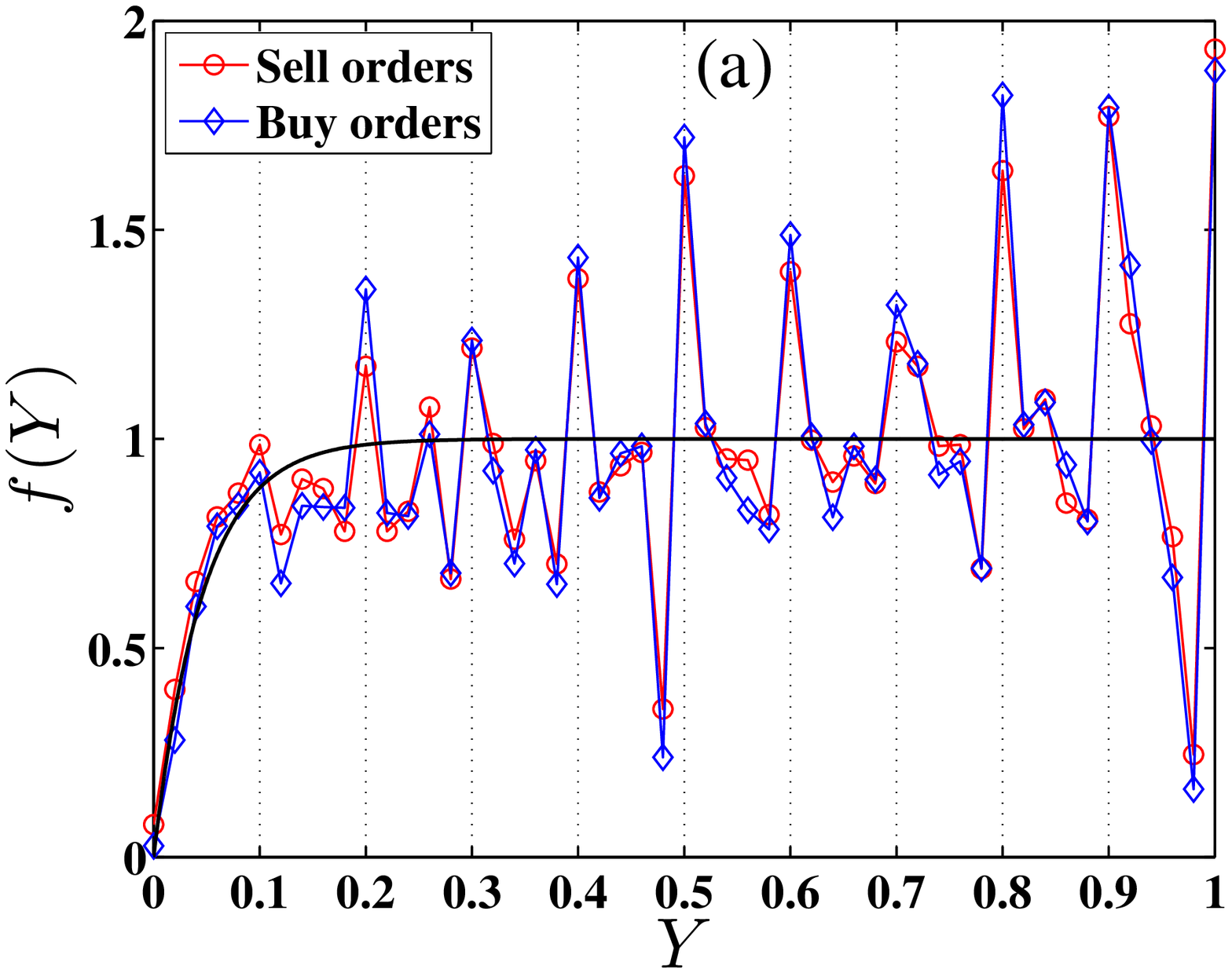}
\includegraphics[width=6cm]{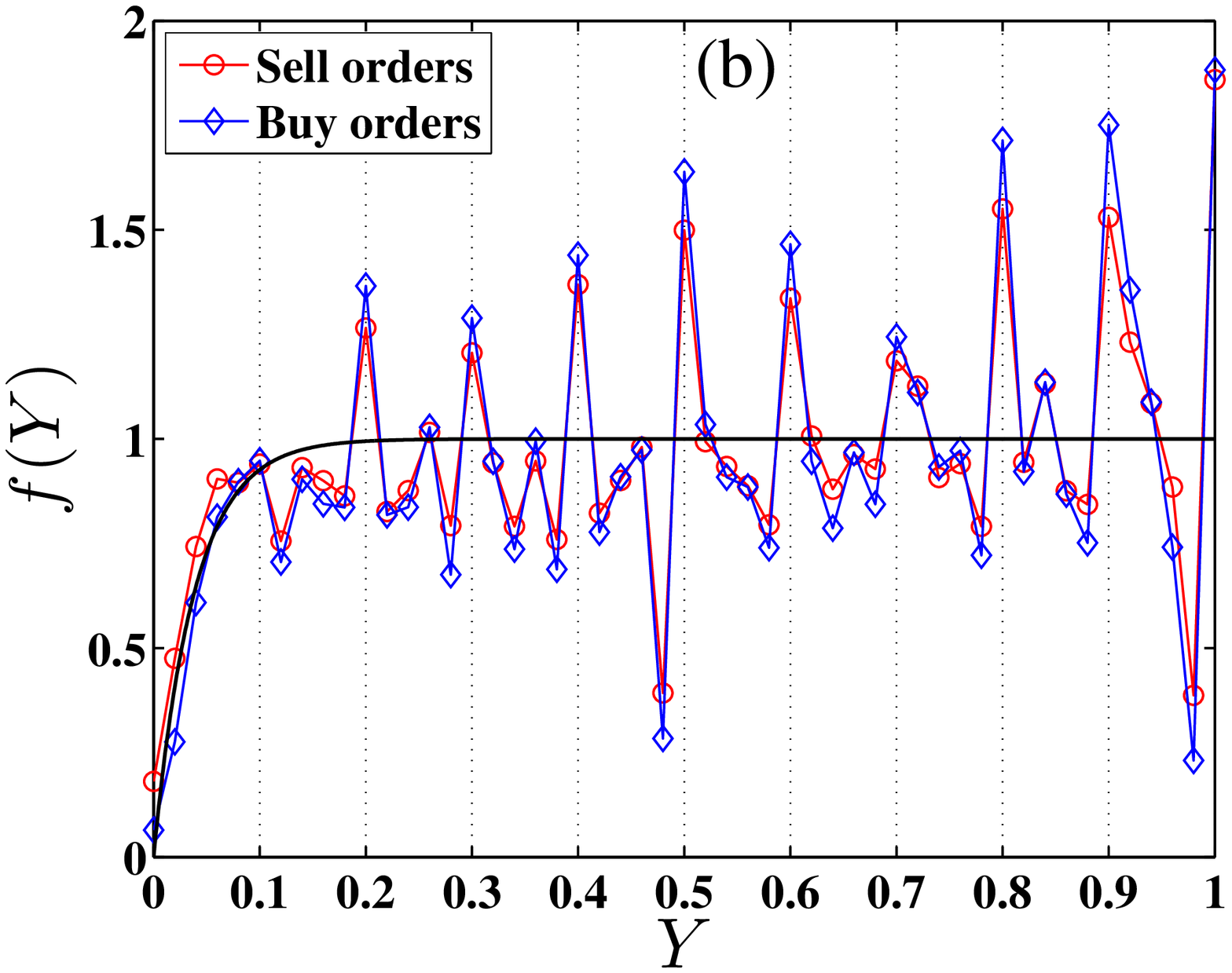}
\includegraphics[width=6cm]{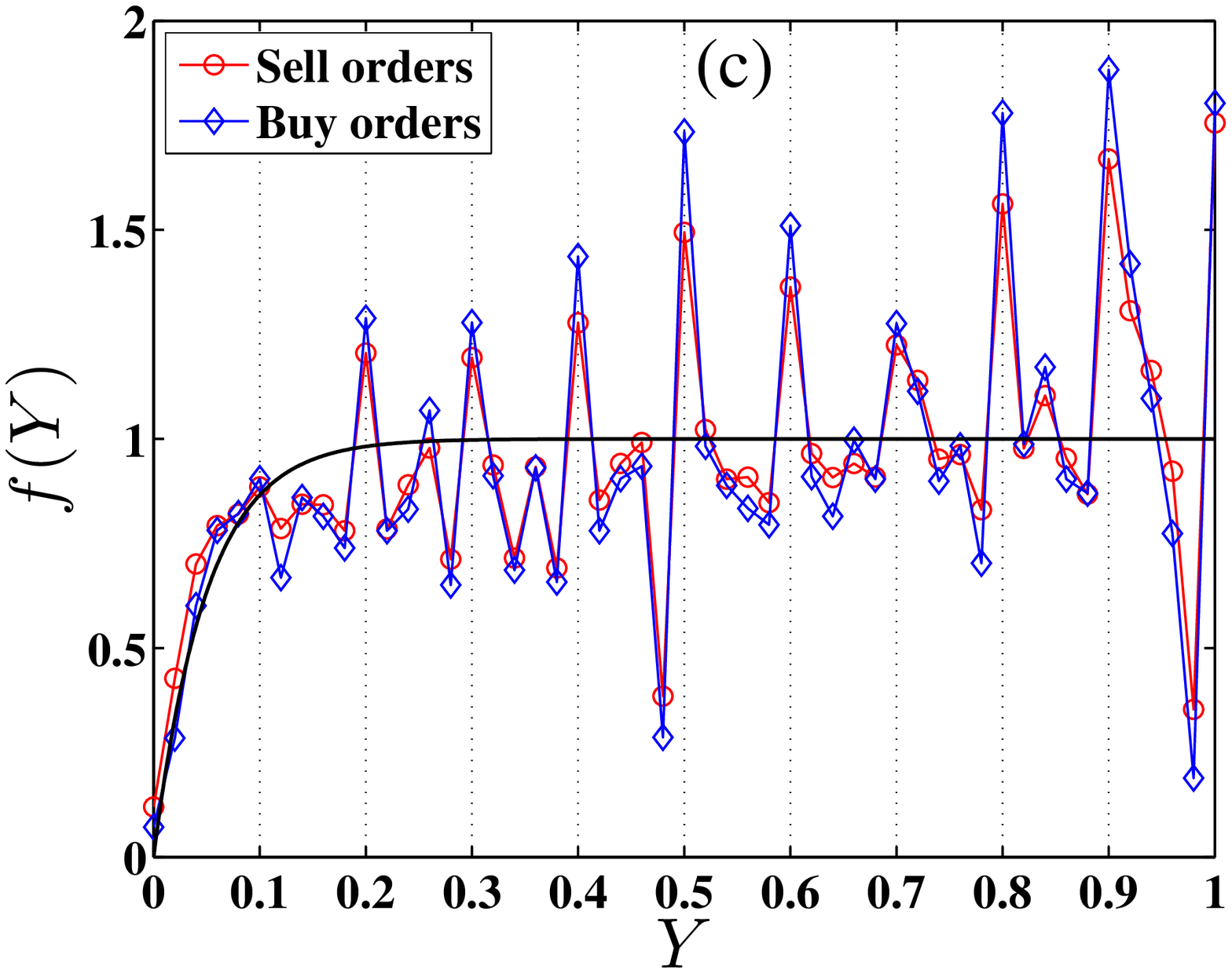}
\includegraphics[width=6cm]{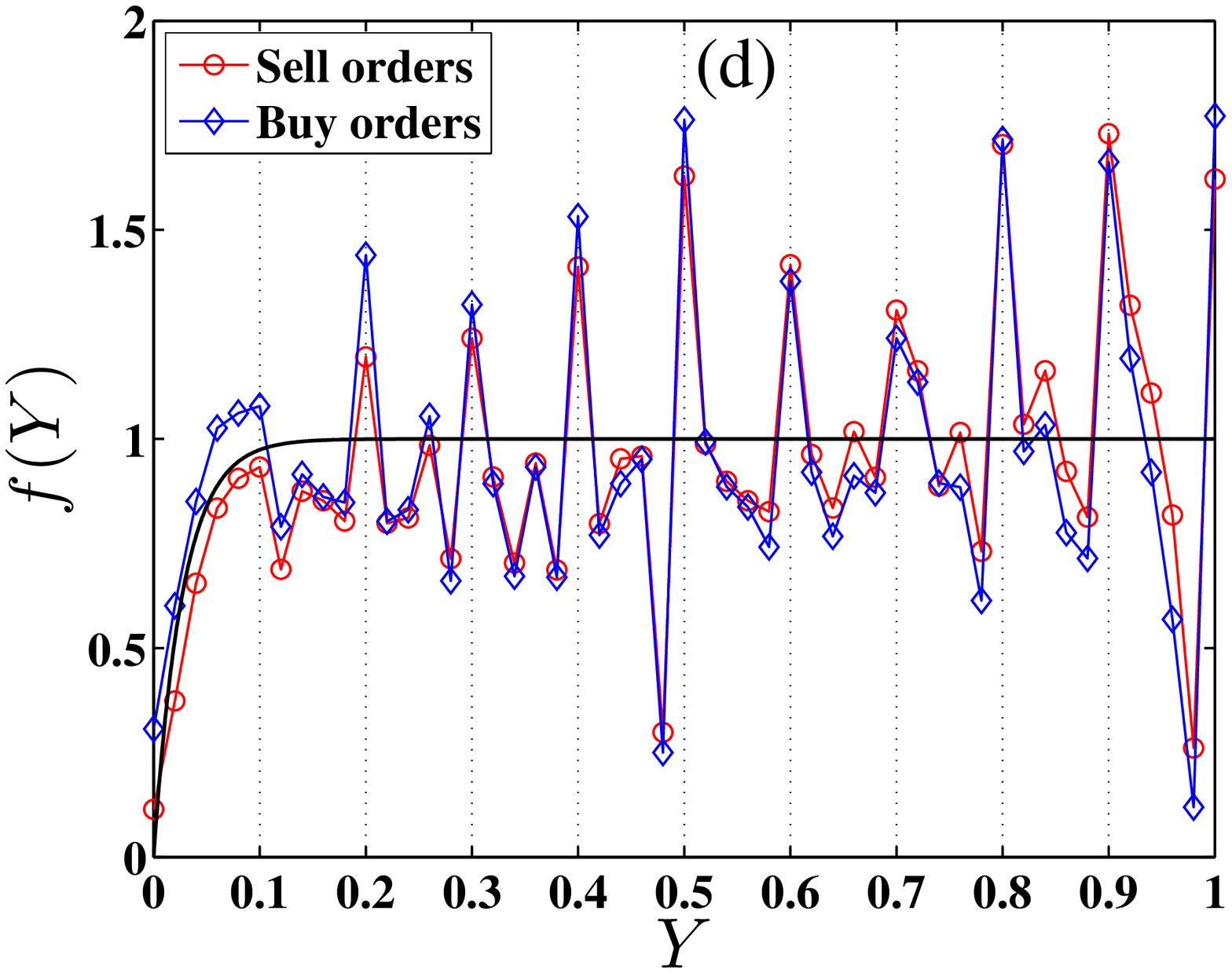}
\caption{Probability density functions $f(Y)$ of aggregating relative positions for both buy and sell orders of four stocks, 000012 (a), 000021 (b), 000066 (c) and 000625 (d). The solid curves are the best fitting function illustrated in Eq.~(\ref{Eq:fYz}).}
\label{Fig:ps_pl_stock}
\end{figure}

We also analyze the PDFs of relative positions for a randomly chosen price levels and find that it has a similar distribution. So we aggregate the data $Y(x,t)$ at all price levels together, and calculate the ensemble PDFs of relative positions. Figure~\ref{Fig:ps_pl_stock} presents the PDFs $f(Y)$ of four stocks randomly chosen from the 23 stocks. We find that the aggregating PDF is similar to the PDF at a certain price level illustrated in figure \ref{Fig:ps_pl_pl}. In addition, we estimate the exponents $\beta$ for the rest stocks, and the values are illustrated in table~\ref{Tb:FitLN}.

\begin{table}[htp]
\centering
\caption{Characteristic parameters of PDFs for both buy and sell orders of 23 stocks. The parameters $\beta$ and $rms$ have the same meaning mentioned in table \ref{Tb:Lognormal-pl}.}
\medskip
\label{Tb:FitLN}
\centering
\begin{tabular}{crrrrc|ccrrrr}
 \hline \hline
 \multirow{3}*[2mm]{Stock} & \multicolumn{2}{c}{Buy orders} & \multicolumn{2}{c}{Sell orders} && \multirow{3}*[2mm]{Stock} & \multicolumn{2}{c}{Buy orders} & \multicolumn{2}{c}{Sell orders} \\
 \cline{2-5}\cline{8-11}
 & \multicolumn{1}{c}{$\beta$} & \multicolumn{1}{c}{$rms$} & \multicolumn{1}{c}{$\beta$} & \multicolumn{1}{c}{$rms$} &&& \multicolumn{1}{c}{$\beta$} & \multicolumn{1}{c}{$rms$} & \multicolumn{1}{c}{$\beta$} & \multicolumn{1}{c}{$rms$} \\
  \hline
    000001 & -33.78 & 0.17 & -36.57 & 0.16 && 000406 & -15.12 & 0.29 & -18.94 & 0.24 \\
    000002 &  -9.42 & 0.29 & -14.35 & 0.22 && 000429 & -10.15 & 0.36 & -12.42 & 0.28 \\
    000009 & -14.07 & 0.22 & -18.90 & 0.19 && 000488 & -23.40 & 0.41 & -12.32 & 0.37 \\
    000012 & -18.81 & 0.31 & -24.12 & 0.27 && 000539 & -29.23 & 0.70 & -28.82 & 0.31 \\
    000016 & -12.72 & 0.36 & -18.21 & 0.30 && 000541 &  -9.55 & 0.43 & -16.41 & 0.39 \\
    000021 & -21.26 & 0.29 & -30.91 & 0.23 && 000550 & -35.81 & 0.30 & -25.01 & 0.25 \\
    000024 & -17.67 & 0.41 & -17.23 & 0.32 && 000581 & -30.32 & 0.39 & -14.85 & 0.33 \\
    000027 & -13.00 & 0.32 & -13.04 & 0.23 && 000625 & -40.17 & 0.32 & -22.77 & 0.28 \\
    000063 & -44.70 & 0.36 & -33.26 & 0.24 && 000709 & -13.25 & 0.26 & -12.91 & 0.24 \\
    000066 & -17.81 & 0.32 & -22.36 & 0.24 && 000720 & -28.33 & 1.10 & -34.40 & 0.86 \\
    000088 & -32.43 & 0.51 & -21.70 & 0.46 && 000778 & -12.38 & 0.35 & -17.15 & 0.30 \\
    000089 & -10.89 & 0.41 & -14.46 & 0.31 && \\
  \hline\hline
 \end{tabular}
\end{table}

Moreover, an interesting feature has been observed that the PDF $f(Y)$ has periodic peaks at $Y=0.1m$ ($m=1,~2,\cdots,~10$) for both buy and sell orders. This pattern is rather robust if we use other bins. This periodic strip pattern is also observed in the snapshot of stocks traded on the London Stock Exchange \cite{Mike-Farmer-2008-JEDC}, in the LOB shape in the Chinese stock market \cite{Gu-Chen-Zhou-2008c-PA}, and in the PDF of order sizes in the opening call auction in the Chinese stock market \cite{Gu-Ren-Ni-Chen-Zhou-2010-PA}. The underlying mechanisms of these periodic behaviors might be different, most of which are still unknown. For $f(Y)$, the mechanism is rather mechanical and stems from the discreteness of the length $n_{bs}$ of the queue. The two largest peaks at $Y=0.5$ and $Y=1$ can be easily explained. Assume that any position $y$ has the same cancellation probability for a given queue. For any queue of length $n_{bs}$ in the limit order book, the occurrence probability of $Y=1$ is $P(Y=1)=1/n_{bs}$. For a queue of even length, we have $P(Y=0.5)=1/(2n_{bs})$. When we average the probability over many queues with different lengths, the values of these two $Y$'s stand out.

\begin{figure}[htb]
\centering
\includegraphics[width=6cm]{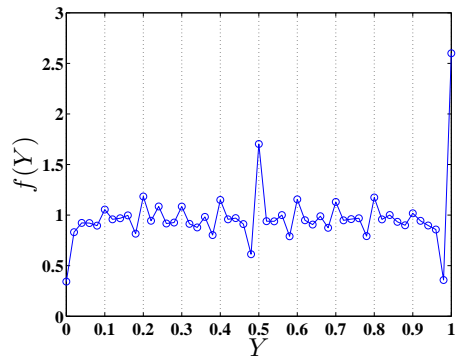}
\caption{Probability density functions $f(Y)$ obtained from the numerical experiment.}
\label{Fig:Simulation}
\end{figure}

In the general case of all peaks, the interpretation is not straightforward. We find that the histogram shape of $n_{bs}$ is quite similar to the order book shape \cite{Gu-Chen-Zhou-2008c-PA}: It increases to a peak around $n_{bs}=10$ and then decays. The relatively large occurrences of $n_{bs}=10$ already imply the possible appearance of peaks at $Y=0.1m$ in $f(Y)$. To test the hypothesis that the discreteness of the queue length causes those peaks, we perform a simple simulation. We assume that $n_{bs}$ is uniformly distributed in [1,100]. For each queue, the cancellation probability at each position is the same. We have generated $10^6$ queues. The resulting $f(Y)$ is illustrated in figure \ref{Fig:Simulation}. In this trivial numerical experiment, we also observe peaks at $Y=0.1m$ with $m=1,~2,\cdots,~10$. This is direct evidence supporting our explanation.

\section{Conclusion}
 \label{sec:conclusion}

Cancellation plays an important role in the dynamics of price formation in the financial markets and helps traders to avoid NE risk or FO risk. In this paper, we have analyzed the order flow data of 23 liquid stocks traded on the Shenzhen Stock Exchange in the year 2003 and studied the empirical the position profiles of order cancellations by rebuilding the buy and sell limit-order books. The main aim of this work is to provide information about order cancellations, which is expected to be used in the construction of zero-intelligence order-driven models \cite{Mike-Farmer-2008-JEDC,Gu-Zhou-2009-EPL}.

We first analyzed the position profiles of relative price levels where cancellations allocate and find that the probability density function (PDF) obeys the log-normal distribution. It implies that when impatient traders can not make an transaction immediately, they will cancel the placed orders and place new orders to catch the price in order to avoid NE risk. Moreover, we observed that the PDF of cancelled sell orders has a higher peak near the same best price than the buy cancellations. The result is consistent with the fact that the Chinese stock market was bear from 2001 to 2005 and traders were more inclined to submit sell orders. We further studied the PDF of cancellation positions for the normalized relative price levels, and found that the PDF follows a power-law behavior in the tail. We also studied the PDF of cancellation positions at a certain price level, and found that the PDF increases rapidly in the front of the queue and then fluctuates around a constat value until the end of the LOB. In addition, the PDF can be fitted by exponent functions for both buy and sell cancellations. It means that in the Chinese stock market, orders earlier submitted are less to be cancelled, which may be caused by the reason that most traders are patient and FO risk is not the main reason of cancellation.

Further work can be conducted to analyze the statistical properties of lifetime of cancelled orders and to study the relation among cancellation position, order lifetime, order size and so on. It will be certainly interesting.

\section*{Acknowledgments}

This work was partly supported by National Natural Science Foundation of China (Grants No. 71101052, 71131007 and 11075054), Shanghai ``Chen Guang'' Project (Grant No. 2010CG32), Shanghai Rising Star (Follow-up) Program (Grant No. 11QH1400800), and the Fundamental Research Funds for the Central Universities.

\section*{References}

\bibliography{E:/Papers/Auxiliary/Bibliography}

\end{document}